\documentclass[11pt,preprint]{aastex}
\usepackage{color}
\usepackage{ulem}

\slugcomment{draft version \today, }

\shorttitle{The influence of inelastic neutrino reactions with light nuclei on the SASI in core-collapse supernovae}
\shortauthors{Furusawa et al.}

\begin{document}

\title{The influence of inelastic neutrino reactions with light nuclei on the standing accretion shock instability in core-collapse supernovae}

\author{Shun Furusawa\altaffilmark{1}, Hiroki Nagakura\altaffilmark{1,2}, Kohsuke Sumiyoshi\altaffilmark{3} and Shoichi Yamada\altaffilmark{1,4} }

\altaffiltext{1}{Advanced Research Institute for Science and Engineering, Waseda University, 3-4-1, Okubo, Shinjuku, Tokyo 169-8555, Japan}
\altaffiltext{2}{Yukawa Institute for Theoretical Physics, Kyoto University, Oiwake-cho, Kitashirakawa, Sakyo-ku, Kyoto, 606-8502, Japan}
\altaffiltext{3}{Numazu College of Technology, Ooka 3600, Numazu, Shizuoka 410-8501, Japan}
\altaffiltext{4}{Department of Science and Engineering, Waseda University, 3-4-1 Okubo, Shinjuku, Tokyo 169-8555, Japan}
\email{furusawa@heap.phys.waseda.ac.jp}

\begin{abstract}
We perform numerical experiments to investigate
 the influence of inelastic neutrino reactions with light nuclei on
 the standing accretion shock instability (SASI).
The time evolution of shock waves is calculated 
with a simple light-bulb approximation for the neutrino transport 
and a multi-nuclei equation of state. 
The neutrino absorptions and inelastic interactions with deuterons, tritons, helions and alpha particles are taken into account
 in the hydrodynamical simulations in addition to the ordinary charged-current interactions with nucleons. 
Axial symmetry is assumed but no equatorial symmetry is imposed.
We show that the heating rates of deuterons reach as high as $\sim 10  \%$ of those of nucleons around the bottom of the gain region.
On the other hand, alpha particles are heated  near the shock wave,
which is important when the shock wave expands and the density and temperature of matter become low.
It is also found that the models with heating by light nuclei have different evolutions from those without it in the non-linear phase of SASI.
This results is because matter in the gain region has a varying density and temperature and there appear sub-regions that are locally rich in deuterons and alpha particles. 
Although the light nuclei are never  dominant heating sources and they work favorably for shock revival in some cases and unfavorably in other cases, they are non-negligible and warrant further investigation.
\end{abstract}
\section{Introduction}
The mechanism of core-collapse supernovae is  not clearly understood at present because of its intricacy (see e.g. \citet{Kotake06,Janka12,Burrows12}).
Many numerical simulations performed by different groups have consistently demonstrated that 
the shock waves formed by the bounce of collapsing cores are
 decelerated and stalled by the energy losses due to the dissociations of nuclei and emissions of neutrinos.
For the moment, the neutrino-heating is considered
 to be the most promising mechanism of shock revival,
 in which the neutrinos emitted from the proto-neutron star reinvigorate the stalled shock to propagate outward again,
 although some other mechanisms, e.g. magneto-rotational explosion, may be needed for very massive stars.
It is also believed that the so-called standing accretion shock instability (SASI)
and  convection 
are essential to increase the efficiency of neutrino heating
\citep{Herant92,Burrows95,Fryer02, Blondin03,Fryer04,Scheck04,Scheck06,Ohnishi06,Foglizzo06,Foglizzo07,Foglizzo12,Iwakami08,Iwakami09,
Fernandez09a,Fernandez09b,Fernandez10,Hanke12,Hanke13, Muller12, Bruenn13, Ott13, Murphy13}.
Indeed, it is recently found that multi-D numerical simulations 
have successfully relaunched the stalled shock wave, which may eventually produce supernova explosions as we see them
\citep{Buras06a,Buras06b,Marek09a,Suwa10,Suwa11,Muller12,Kuroda12,Takiwaki12,Bruenn13, Ott13,Murphy13}.
All these simulations are not long enough so far, however, and it is remaining to see if they can really reproduce the canonical explosion energy and $^{56}$Ni mass \citep{Yamamoto13}. 
In addition to these hydrodynamical effects, there are  some nuclear-physical ingredients
 that are also supposed to be important for reproducing the core-collapse supernovae.
Nuclear burning in the accreting matter and ejecta was investigated by \citet{Nakamura12} and \citet{Yamamoto13}.
The inelastic  neutrino interactions as well as the baryonic equation of state (EOS) are also important as described below.
 
 The inelastic interactions between neutrinos and nuclei have been neglected
in most hydrodynamical simulations of the neutrino heating.
\citet{Haxton88} was the first to point out the importance of these reactions.
\citet{OConnor07}, \citet{Arcones08}, \cite{Langanke08} and \citet{Barnea08} 
investigated their  influences on the dynamics, neutrino spectrum as well as nucleosynthesis.
There is, however, no investigation of the impact of the inelastic reactions on multi-D hydrodynamics but \citet{Ohnishi07}.
They showed that the inelastic neutrino interactions with alpha particles are helpful to revive the shock in 2D  simulations 
if the neutrino luminosity is close to the critical value, which is the threshold for a shock revival.
They took into account only alpha particles as additional heating sources, since the mass fractions 
of other nuclei  were not available in the EOS they used \citep{Shen98a,Shen98b}. 
However, the shocked matter is  certainly composed  not only of nucleons and
 alpha particles but also of deuterons, tritons and helions \citep{Sumiyoshi08,Arcones08,Hempel12}.
The energy-transfer coefficients, that is, the average values of the product of the cross 
section and energy transfer of deuterons are comparable to those of nucleons
and ten times greater than those of alpha particles \citep{Nakamura09}.
Tritons and helions have also larger energy-transfer coefficients than alpha particles \citep{OConnor07,Arcones08,Nakamura09}.

The EOS is another important physical input in supernova simulations and its influences on the dynamics of core-collapse supernovae has been investigated
 by many researchers, e.g. \citet{Sumiyoshi05}, \citet{Marek09b}, \citet{Hempel12}, \citet{Suwa13} and \citet{Couch13} to mention a few.
There are currently two EOS's that are widely used for the simulations: Lattimer-Swesty's EOS \citep{Lattimer91} and Shen's EOS \citep{Shen98a,Shen98b,Shen11}. 
In both EOS's, the ensemble of heavy and light nuclei is 
approximated by a single representative heavy nucleus and alpha particle.
In this decade, however, some EOS's that incorporate a large number of nuclei have been constructed \citep{Botvina04,Botvina10, Hempel10, Blinnikov11, G.Shen11}.
We have also developed such an EOS \citep{Furusawa11,Furusawa13}.
We employ the liquid drop model for heavy nuclei and take into account shell effects and nuclear pasta phases.
Unbound nucleons are treated by the relativistic mean field theory \citep{Furusawa11}.
Moreover, we implement
some important improvements such as  the inclusion of the Pauli and self-energy shifts in 
the mass evaluation of light nuclei \citep{Furusawa13}.
As a result, the mass fractions of various light nuclei have become much more reliable.

The aim of this paper is to investigate the impacts of the inelastic neutrino reactions with light nuclei
on the SASI based on our new EOS.
We perform experimental simulations of the post-bounce phase in 2D,
 employing the light bulb approximation for neutrino transfer.
In addition to the ordinary cooling and heating by nucleons, we incorporate the heating by deuterons,
tritons, helions and alpha particles.
This article is organized as follows. 
In section 2, we describe some important ingredients in numerical simulations such as
the hydrodynamics code and  the rates of  inelastic reactions with light nuclei that we employ in this study.
Then the results are shown in section 3, with an emphasis being put on the role of light nuclei in the shock heating. 
The paper is wrapped up with a summary and some discussions in section~4.

\section{Models}
The basic set-up of our dynamical simulations is the same as that given in \citet{Ohnishi06,Ohnishi07} and \citet{Nagakura13}
except for the inelastic reactions with light nuclei.
We perform 2D simulations assuming axial symmetry.
Spherical coordinates are used and no equatorial symmetry is assumed. 
We utilize 300 radial mesh points to cover $r_{in} \leq r \leq r_{out} \ (=500 \ \rm{km})$,
 where $r_{in}$, the inner boundary of the computation domain,  is set to be the radius of the neutrino sphere of $\nu_{e}$ in the initial state. 
We deploy 60 angular mesh points to cover the whole meridian section.
We solve the following equations
\begin{equation}
 \frac{d\rho}{dt} + \rho \nabla \cdot \mbox{\boldmath$v$} = 0,
\end{equation}
\begin{equation}
 \rho \frac{d \mbox{\boldmath$v$}}{dt} = - \nabla  p  + \rho\nabla \Bigl( \frac{G M_{\rm in}}{r} \Bigr),
\end{equation}
\begin{equation}
 \rho \frac{d}{dt}\displaystyle{\Bigl(\frac{e}{\rho}\Bigr)}  = - p 
\nabla \cdot 
\mbox{\boldmath$v$} +
 Q_{E}
  + Q_{d}+ Q_{t}+ Q_{h}+ Q_{\alpha},
  \label{eq:energy}
\end{equation}
\begin{equation}
 \frac{dY_{e}}{dt} = Q_{N},
  \label{eq:yeevo}
\end{equation}
where $\rho$, $p$, $T$, $e$ and $Y_{e}$ denote the mass density, pressure, temperature, energy density 
and electron fraction, respectively. Other symbols, $r$, $\mbox{\boldmath$v$}$, and $G$, stand for the radius, fluid velocity and 
gravitational constant, respectively. 
 The mass of a central object, $M_{\rm{in}}$, is assumed to be constant and set to be $M_{\rm{in}}=1.4 M_{\odot}$.
Interactions between neutrinos and nucleons are encapsulated in $Q_{E}$ and $Q_{N}$,
the expressions of which are adopted from Eqs. (16) and (17) in \citet{Ohnishi06}.
$Q_{d,t,h,\alpha}$ are  the heating rates for the light nuclei indicated by the subscripts.
The heating for alpha particles, $Q_{\alpha}$,  corresponds to  $Q_{\rm{inel}}$ in Eqs.(3) and (6) in \citet{Ohnishi07},
where only alpha particles were taken into account as the additional heating source. 
The heating rates for deuterons, tritons and helions, $Q_{d,t,h}$, are the new elements  in this work.

The neutrino transport is handled by the simple light bulb approximation,
in which neutrinos with Fermi-Dirac distributions are assumed to be emitted from the proto-neutron star and travel radially.
We also assume that the temperatures of $\nu_{\rm{e}}$, $\bar{\nu_{\rm{e}}}$ and $\nu_{\rm{\mu}}$  are constant
and set to be  $(T_{\nu_{\rm{e}}},T_{\bar{\nu}_{\rm{e}}},T_{\nu_{\rm{\mu}}})=(4,5,10)$ in MeV.
The luminosities of  $\nu_{\rm{e}}$ and $\bar{\nu_{\rm{e}}}$  are assumed to have the same value: $L_{\nu_{\rm{e}}}=L_{\bar{\nu}_{\rm{e}}}=L$.
The luminosity of $\nu_{\rm{\mu}}$ is set to be  half that value, $L_{\nu_{\rm{\mu}}}=0.5 \times L$,  as in \citet{Ohnishi07}.
The numerical code for hydrodynamics is based on the central scheme, which is a popular choice at present (see, e.g., \citet{Nagakura08,Nagakura11})
and the implementation of the light bulb approximation is explained in detail in \citet{Nagakura13}.
We employ the multi-nuclei EOS, which gives not only thermodynamical quantities but also the abundance of various light and heavy nuclei
up to the mass number $A \sim 1000$ \citep{Furusawa11,Furusawa13}.

The heating rates for light nuclei are calculated from the analytic formula given in \citet{Haxton88}. 
\begin{eqnarray}
 Q_{i}  = \frac{\rho X_{i}}{m_{u}}  \frac{31.6 \ \rm{MeV \ s^{-1}}}{(r/10^{7} \rm{cm})^{2}}
  &&\left[
     \frac{L_{\nu_{\rm{e}}}}{10^{52}\rm{ergs~s}^{-1}}
     \left(\frac{5\rm{MeV}}{T_{\nu_{\rm{e}}}}\right)
     \frac{A_i^{-1}\langle
     \sigma_{\nu_{\rm{e}}}^{+}E_{\nu_{\rm{e}}}
     +\sigma_{\nu_{\rm{e}}}^{0}E_{\rm{NC}}^{i}
     \rangle_{T_{\nu_{\rm{e}}}}}
     {10^{-40}\rm{cm}^2\rm{MeV}} \right. \nonumber \\
 &&+\quad\frac{L_{\bar{\nu}_{\rm{e}}}}{10^{52}\rm{ergs~s}^{-1}}
  \left(\frac{5\rm{MeV}}{T_{\bar{\nu}_{\rm{e}}}}\right)
  \frac{A_i^{-1}\langle
  \sigma_{\bar{\nu}_{\rm{e}}}^{-}E_{\nu_{\rm{e}}}
  +\sigma_{\bar{\nu}_{\rm{e}}}^{0}E_{\rm{NC}}^{i}
  \rangle_{T_{\bar{\nu}_{\rm{e}}}}}
  {10^{-40}\rm{cm}^2\rm{MeV}} \nonumber \\
 &&+\left.\quad
    \frac{L_{\nu_{\mu}}}{10^{52}\rm{ergs~s}^{-1}}
    \left(\frac{10\rm{MeV}}{T_{\nu_{\mu}}}\right)
    \frac{A_i^{-1}\langle
    \sigma_{\nu_{\mu}}^{0}E_{\rm{NC}}^{i}
    +\sigma_{\bar{\nu}_{\mu}}^{0}E_{\rm{NC}}^{i}
    \rangle_{T_{\nu_{\mu}}}}
    {10^{-40}\rm{cm}^2\rm{MeV}}
   \right],
 \label{eq:light}
\end{eqnarray}
where $i$ specifies a light nucleus, $d,t,h$ or $\alpha$;
$A_i$ and $X_{i}$ are the mass number and mass fraction of nucleus $i$, respectively;
$m_{u}$ is the atomic mass unit;
the average over the neutrino spectrum is denoted as $\langle$ $\rangle_T$.
The energy-transfer coefficients for deuterons are calculated from Table~I in \citet{Nakamura09} 
for both the neutral-current (NC), 
$\langle\sigma_{\nu}^{0}E_{\rm{NC}}^{i} \rangle_T$, and
 the charged-current (CC),$\langle\sigma_{\nu}^{\pm}E_{\nu} \rangle_T$.
The energy-transfer coefficients for the nuclei with $A_i=3$ (tritons and helions) are obtained from 
Table~II in \citet{OConnor07} for NC.
As for the CC interactions between tritons and $\bar{\nu}_{e}$, we utilize Table~I in \citet{Arcones08}.
The other CC reactions involving tritons  or helions are not included in our simulations,
 since neither the energy-transfer coefficients nor the cross sections are available.
The effects of helions are negligible, however,
 since the abundance of helions is much smaller than nucleons and dominant light nuclei ($d$ and $\alpha$). 
The energy-transfer coefficients for alpha particles are derived from Table~II in \citet{Haxton88} for CC
  whereas we utilize for NC the fitting formula provided by \cite{Haxton88},
\begin{equation}
 A_i^{-1}\langle
  \sigma_{\nu}^{0}E_{\rm{NC}}^{i}
  +\sigma_{\bar{\nu}}^{0}E_{\rm{NC}}^{i}
  \rangle_{T_{\nu}}
  = \alpha\left[ \frac{T_{\nu} - T_{0}}{10\rm{MeV}} \right]^{\beta},
  \label{eq:fitting}
\end{equation}
where the parameters are chosen to be $\alpha=1.28\times 10^{-40}$~MeV~cm$^{2}$, $\beta=4.46$
and $T_{0}=2.05$~MeV following \citet{Gazit04}.
The cooling reactions involving light nuclei are ignored, since the reaction rates are not available at the moment.
 In this sense, the influences of light nuclei that we find in this paper should be regarded as the maximum.
We also ignore the contributions of charged-current interactions of light nuclei to the evolution of electron fraction Eq.~(\ref{eq:yeevo}) as in \citet{Ohnishi07},
since they are quite minor compared to the contributions of nucleons in most regions as demonstrated at  the end of the next section.

As the first step of the simulations, we prepare the initial conditions,
 which are spherically symmetric, steady accretion flows
 that are stable to radial perturbations \citep{Yamasaki05,Ohnishi06,Nagakura13}.
The inelastic interactions of neutrinos with light nuclei, $Q_{d,t,h,\alpha}$, are also included in these computations.
We start dynamical simulations, adding radial-velocity perturbations of 1 $\%$,
 which are proportional to $\rm{cos}\theta$.
We vary the luminosity $L$ and mass accretion rate $\dot{M}$ 
and investigate the influences of light nuclei on dynamics under different physical conditions.
We refer to  the normalized neutrino luminosity $L_{52}\equiv L /(10^{52} \rm{erg/sec})$ 
and mass accretion rate $\dot{M}_{sun} \equiv -\dot{M}/(M_{\odot} \rm{s^{-1}})$ in specifying models.
\section{Result} 
In the following subsections we first discuss the influences of light nuclei on the initial states,
 that is, the spherically symmetric, steady accretion flows
 through the standing shock wave onto the proto-neutron star.
Then, a 1D simulation is presented to give an insight into the roles of light nuclei. 
Finally, we describe the results of  2D dynamical simulations in detail.

In Table 1, we compare the heating rates per baryon for light nuclei,
which can be evaluated without referring to matter profiles.
 We set $r =100$ km, $L_{52}=5.0$ and $X_i = 1.0$.
Note that the cooling rates are not subtracted here for comparison.
We set $X_p = X_n = 0.5$ in the evaluations of the heating rates for nucleons.
It is found that deuterons have the heating rates per baryon that are comparable to those of nucleons.
Tritons, helions and alpha particles have rather small heating rates per baryon.  
Muon neutrinos do not heat nucleons but light nuclei thorough NC,
 since the former has no internal degree of freedom that can be excited.

\subsection{Steady state \label{steady}}
\begin{figure}
\begin{center}   
\begin{tabular}{ll}
  \resizebox{58mm}{!}{\plotone{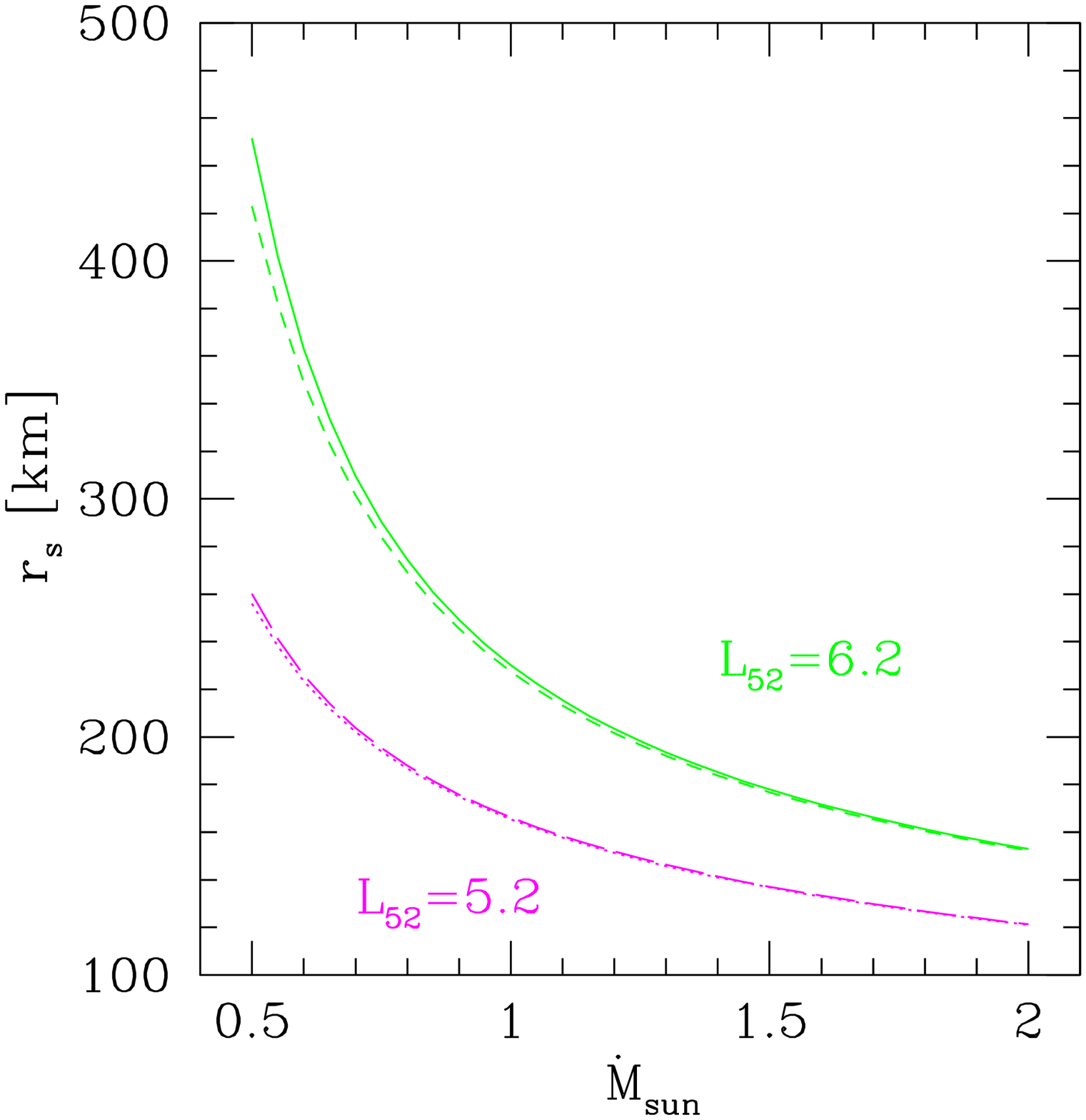}}   \resizebox{52mm}{!}{\plotone{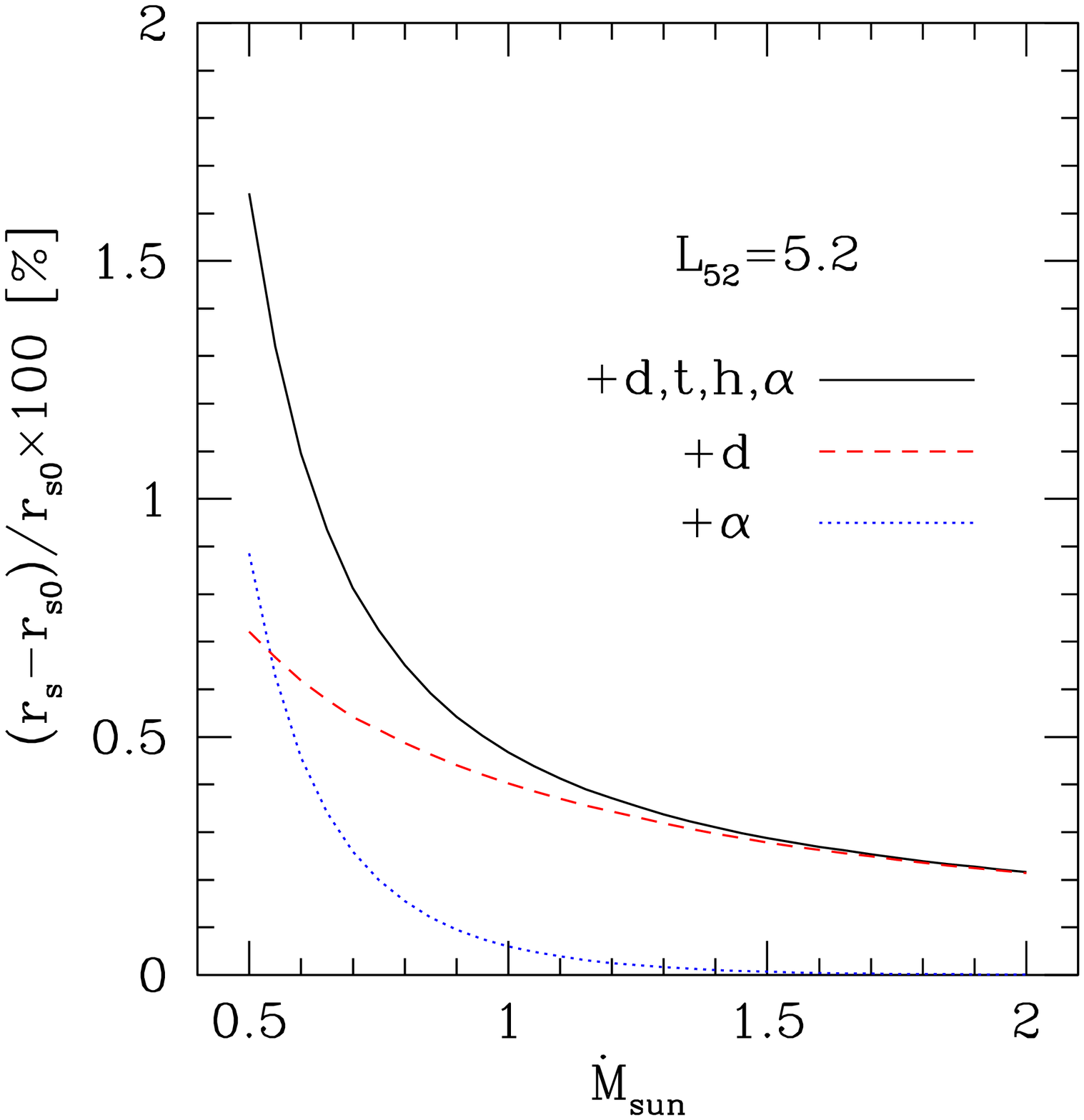}}   \resizebox{52mm}{!}{\plotone{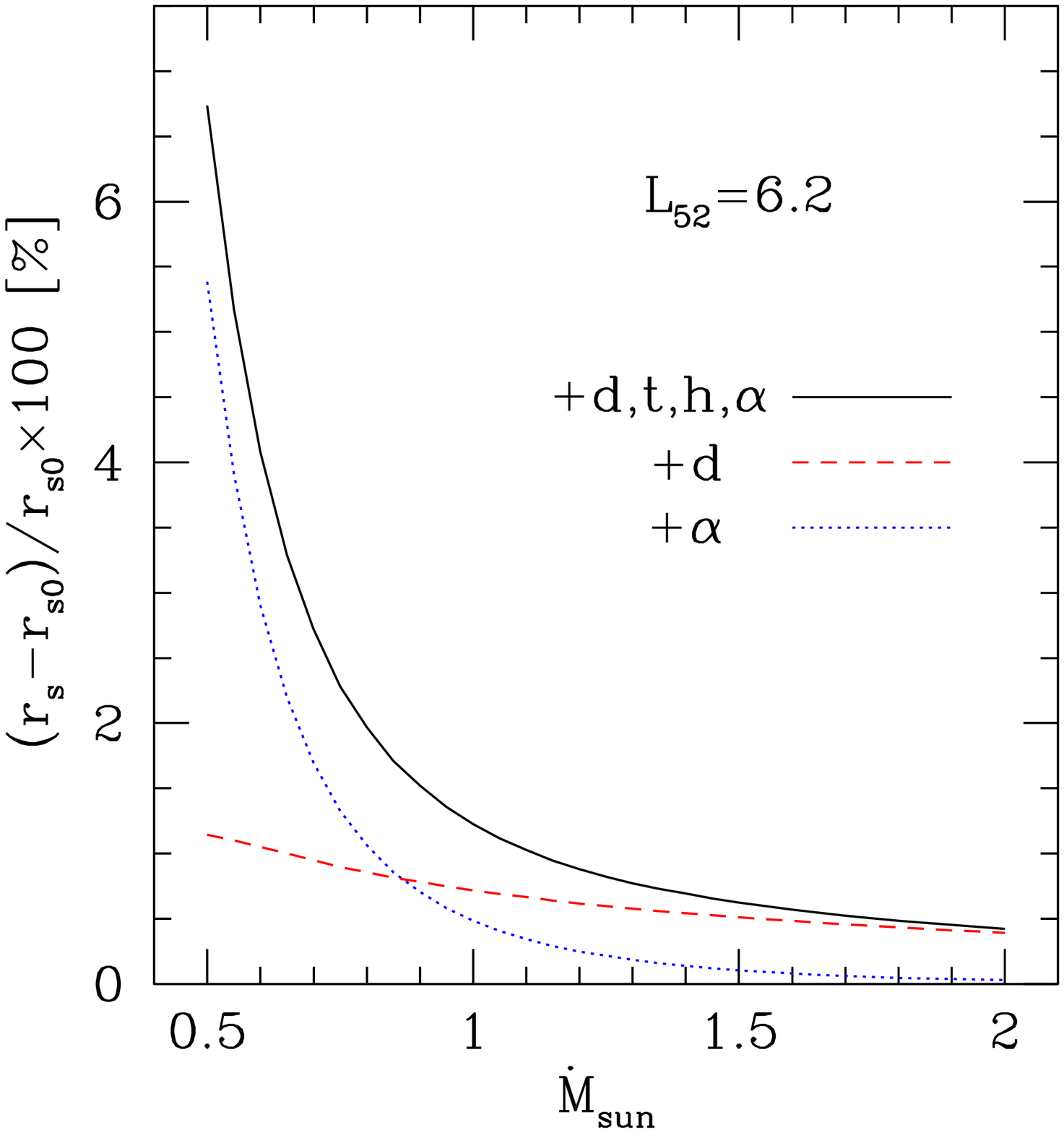}} \\
    \end{tabular}
\end{center}
\caption{The left panel shows the shock radii in the initial steady states for $L_{52}=6.2$
 with the light-nuclei heating (green solid line) and without it (green dashed line)
and for $L_{52}=5.2$ with the light-nuclei heating (magenta long dashed line) and without it (magenta doted line).
The center and right panels show the variations of shock radii due to the heating reactions
 with all light nuclei (black solid line), only deuterons (red dashed line) and only alpha particles (blue dot line).}
\label{drst}
\end{figure}

The left panel of Fig.~\ref{drst}  displays the shock radii $r_s$ in the spherically symmetric, steady accretion flows 
with and without the heating of light nuclei for $L_{52}=5.2$ and 6.2.
We can see the shock radii are not significantly affected by this change.
Higher luminosities and lower mass accretion rates make the light nuclei a little bit more influential on the structures of the steady states.   
Note also that higher luminosities and lower mass accretion rates
 result in the steady states that are closer to the critical ones and
 small variations may have a greater effect.
The center and right panels of Fig.~\ref{drst} show
 the variations of shock radii, which are defined as $(r_s-r_{s0})/r_{s0}\times100 \ [\%]$
 with the shock radius $r_{s0}$ for no light-nuclei heating, for the three cases,
in which we include either the heating of all light nuclei or that of only deuterons or alpha particles, respectively.
The results indicate that deuterons are
 always one of the main contributors to the heating though the resultant variations are not so large.
On the other hand, alpha particles can push the shock wave only when the mass accretion rate is small. 
The nuclei with $A_i=3$ have little influence in any condition, since tritons and helions are much less abundant than the other two light nuclei.  

Figure~\ref{fraheast} displays the profiles of the mass fraction and heating rate of each nuclear species for two models,
which include  heating by all light nuclei.
For the model with $L_{52}=5.2$ and $\dot{M}_{sun}$= 1.5, the shock radius is $\sim$ $140$ km
 and the deuteron-heating is second dominant after that of nucleons
 in the gain region, that is, the region between the gain radius $r_g$ and shock radius $r_s$.
The gain radius $r_g$ is defined as the radius, at which the neutrino heating is equal to the neutrino cooling
and there is no net energy gain.
 For the model with $L_{52}=6.2$ and $\dot{M}_{sun}$= 0.5, alpha particles contribute to
 the heating as well as nucleons in the outer part of the gain region. 
This difference between deuterons and alpha particles can be also seen
in  Fig.~\ref{lphasest},
which indicates in the $\rho$-$T$ plane under the condition of  $Y_e = 0.5$ the regions, 
 where deuterons and alpha particles are abundant.
 Superimposed are the actual $(\rho, T)$ values obtained in the gain region. 
Note that the electron fractions obtained in the simulations take various values between 0.3 and 0.5 in the gain region.
However, the deuteron-rich and alpha-rich regions for $Y_e=0.3$ are not much different from those for $Y_e = 0.5$.
  
For the model with $L_{52}=6.2$ and $\dot{M}_{sun}=$ 0.5 (brown), 
the shock radius is rather large and the plots of $(\rho, T)$ pairs 
obtained in this model extend to lower densities and temperatures,
 which favor the existence of alpha particles.
For the model with $L_{52}=5.2$ and $\dot{M}_{sun}=$ 1.5 (magenta), on the other hand,
all the ($\rho, T$) pairs are outside the region,
 in which the fraction of alpha particles is larger than 10 $\%$,
 and are located close to the region, in which more than 1 $\%$ of deuteron fractions is realized.
Note that deuterons have the energy-transfer coefficients more than 10 times larger than those of alpha particles
as shown in Tab. 1.
Deuterons can hence make some small contributions to
the total heating rates even if the fraction is as small as $\sim 1 \ \%$. 
Deuterons and alpha particles can contribute to the neutrino heating
 at the inner and outer parts of the gain region, respectively.
As a result, the larger shock radius leads to the more efficient heating by alpha particles than by deuterons.

So far we have investigated the differences that the heating via light nuclei may make
 by arbitrarily switching them on and off for the same background models.
It may be interesting, however, to make comparisons against the models,
 in which not only the heating but also the existence of the light nuclei other than alpha particles is entirely neglected.
 There are two reasons for this: first, the EOS's that have been commonly employed
 in supernova simulations thus far consider only alpha particles as light nuclei; second,
 if light nuclei did not exist in the first place, nucleons would be more populous, taking their places,
 and could be efficiently heated, thus reducing or even nullifying the differences we have observed above.
 
In order to see this, we employ the Shen's EOS (\citet{Shen11}),
 one of the standard EOS's for supernova simulations, in which only alpha particles are
 included as light nuclei, and construct spherically symmetric, steady accretion flows and
 investigate the differences that $d, t$ and $h$ make.
 Strictly speaking, there are some differences between our and Shen's EOS other than the treatment of light nuclei.
 For example, the Shen's EOS takes into account a single representative heavy nuclei
 whereas our EOS handles the ensemble of them.
 This is not so important in the post-shock region of our concern, however,
 since the nucleons and light nuclei are dominant there.
 Even the treatment of alpha particles is different between the two EOS's,
 since we take into account the ambient matter effects in evaluating the mass of alpha particles \citep{Furusawa13}.
 This difference, however, manifests itself only at high densities $\rho~\gg ~10^{11}\rm{g}/\rm{cm}^3$,
 the density at the inner boundary in our models. 

Figure \ref{fraheast} shows that the abundances of nucleons and alpha particles are almost identical
 between the corresponding models. 
As a result, the heating via nucleons does not differ significantly
 although both the mass fractions and heating rates of nucleons and alpha particles are slightly larger
 in the model with the Shen's EOS.
For instance, the mass fractions of nucleons and alpha particles in the model with our EOS
are smaller by 0.9 $\%$ and 0.3 $\%$, respectively, than those in the model with the  Shen's EOS
 at  $r=130$ km for $L_{52}=5.2$ and $\dot{M}_{sun}$= 1.5 
 due to the existence of light nuclei other than alpha particles.
 And the heating rates per baryon ($Q_E$, $Q_{\alpha}$)  
are
(237.0, 0.3643)  in MeV/sec with our EOS,
 which should be compared with the values (238.6, 0.3653) that are obtained with the Shen's EOS.
The total heating rate,  $Q_{E}+Q_{d}+Q_{t}+Q_{h}+Q_{\alpha}$, in the model with our EOS, 
however, is larger than the sum of $Q_E$ and $Q_{\alpha}$
 in the model with the Shen's EOS because of the contribution from deuterons, $Q_{d}=$3.264.
As a result, the shock radius in the former model is slightly larger than that in the latter model.
 The difference is clearer in the case of $L_{52}=6.2$ and $\dot{M}_{sun}=0.5$.
To be fair, we point out that the $Q_E$ includes cooling but others do not completely (see section 2).
If we compare the absorptions of neutrinos alone, the total rates are larger in the models with the Shen's EOS, 
since the contribution from nucleons overwhelms that from deuterons.
Comparisons in dynamical contexts will be given later.

\subsection{1D simulation  \label{sec1D}}
To obtain the basic features of the heating by light nuclei in the dynamical settings,
 we perform a spherically symmetric 1D simulation.
We employ 300 radial mesh points as explained in section 2.
Figure~\ref{timeen} shows the entropy evolution for the model with $L_{52}=$ 5.4 and $\dot{M}_{sun}$= 1.0.
Although we do not add any perturbation initially, numerical noises induce small radial oscillations that grow gradually.
It can be seen that matter just below the shock wave has low entropy
 when the shock radius is large. It should be also noted that the 
entropy for the same shock radius is smaller when the shock is receding than it is proceeding. 
This asymmetric feature becomes clearer as the shock radius gets larger.
In Fig.~\ref{timeev1d}, we compare the time evolutions of the integrated heating rate of each nuclear species
 along with the shock and gain radii.
The shock radius is defined as the iso-entropic surface of  $s=5.0$ $k_B$, where $k_B$ is the Boltzmann constant.
The heating rates are integrated over the gain region as $\int_{\rm{gain}} Q_{i} \rm{d}\bold{r}^3 $  and given in the unit of $10^{52}$ erg~s$^{-1}$.
We can see that the heating rate of alpha particles changes roughly in step with the shock radius,
since the larger shock radius leads to the greater fractions of alpha particles. 
The peak time in the heating rate of alpha particles delays from the time of the local maximum in the shock radius
because of the asymmetric feature in the entropy mentioned above.
It is also found that the heating of alpha particles is more important than that of deuterons after the shock wave revives and goes outward ($t>400$~ms).
On the other hand, the heating rates of deuterons  reach a local maximum when the gain and shock radii are small,
since matter has high densities and favors deuterons.
Furthermore, since deuterons are located closer to the neutrino sphere,
they attain the heating rates as high as 
$1-10 \ \%$ of those of nucleons. 
These results indicate that alpha particles and deuterons heat matter in different phases in the oscillation of the shock wave.
Although tritons and helions are similar to deuterons, they are quite minor.

\subsection{2D simulations}
Figure~\ref{radievo} displays the time evolutions of average shock radii for four models,
in which all light nuclei, only deuterons, only alpha particles and no light nuclei are taken into account in the heating sources, respectively.
The results of the model, in which we employ the Shen's EOS and take into account
 the heating via alpha particles, are also shown and discussed later.
The models without deuteron-heating  do not succeed in the shock revival for $L_{52}=5.1$,
whereas the other two models do though it takes longer times. 
For $L_{52}=5.2$, on the other hand, all models produce shock revival.
We can see that the  heating  by deuterons and alpha particles both reduce the time to shock revival.
The  same trend is also seen in the models with $L_{52}$ = 6.3 and $\dot{M}_{sun} = 1.5$.
These results suggest that the heating of light nuclei, especially deuterons, is helpful for shock revival.
Note, however, that this may be too naive,
since the time to shock revival is known to be sensitive to various ingredients such as the initial perturbations 
when the neutrino luminosity is close to the critical value.
In fact, the models with $L_{52}=6.0$ and $\dot{M}_{sun}=$ 1.5 show the opposite trend
 when we include the deuteron-heating:
the models without the deuteron-heating can gain larger energies through the heating via nucleons alone than other models.
 This is because the deuteron-heating prevents
 the shock wave from shrinking in the first place and 
reduces the heating via nucleons,
 since matter tends to be farther away from the neutrino sphere.
For instance,  at the time  $t=210$ ms,  when all the models
 with $L_{52}=6.0$ and $\dot{M}_{sun}$= 1.5 hit the local minimum in the average shock radius,
 the values of the average shock radius and
 the angle-averaged heating rate per baryon for nucleons are
 178 km and 223 MeV/sec, respectively, in the model without light-nuclei heating,
 whereas they are 183 km and 211 MeV/sec in the model with only deuteron-heating.
 Although we cannot find a clear trend when it is helpful for shock revival,
 the deuteron-heating is non-negligible regardless.
Alpha particles do not work that way, on the other hand. They
heat matter when the shock wave has large radii as discussed in section~\ref{sec1D}
and, as a result,
 do not affect the shock  recession.
The model with the alpha-particle-heating alone has almost the same shock radius 177 km
 at $t=210$ ms. 
 The heating via alpha particles is hence always favorable for shock revival whenever it is effective.

Comparisons between the models with our and Shen's EOS's are a bit more difficult due to the inherent
 intricacies of shock revival mentioned above.
 We can recognize some trends in Fig. 6, however.
 As noted in section 3.1, the heating via nucleons alone is larger in the models with the Shen's EOS
 whereas the total heating rates are greater for the models with our EOS thanks to the contribution from deuterons.
 In accordance with this, shock revival occurs earlier
 in the models that employ our EOS and incorporate the heating via deuterons than in the corresponding models with the Shen's EOS. 
The order is reversed in some cases if the heating of deuterons is switched off in the models with our EOS. 
For example, the stalled shock is revived at $t\sim$ 800~ms for the Shen's EOS in the case of $L_{52}=$5.1;
  for our EOS, it happens at $t\sim600$~ms
 if the heating of deuterons is included whereas the shock remains stalled even at $t \sim \ 1000$~ms if it is turned off. 
In the same way, the model with the Shen's EOS  for $L_{52}=6.0$ shows 
the intermediate time-evolution between the models that employ our EOS with and without the deuteron-heating.
 It should be reminded, however,
 that other differences such as the preheating of nucleons ahead of the shock wave may have some influences on the results. 
 

We now focus on the model with $L_{52}=5.2$ and $\dot{M}_{sun}=1.0$ that includes the heating by all light nuclei 
 to explore in more detail the role of light nuclei in the evolution of the shock wave.
The shock oscillation grows linearly by $t\sim$150~ms in this model as seen in Fig.~\ref{radievo}.
The distributions of nucleons and light nuclei are almost spherically symmetric at $t=$ 100~ms as seen in the upper panels of Fig.~\ref{snap52}.
The heating rates of light nuclei are large in the narrow region near the quasi-steady shock wave at $t=$ 100~ms.
At $t=$ 200 and 300~ms, however, we observe the deformed shock waves that have reached the non-linear regime of SASI.
In some regions, the light nuclei are abundant indeed and their heating is efficient accordingly.
Figure~\ref{lphase52} plots the pairs of $(\rho, T)$ obtained along 5 different radial rays (see Fig.~\ref{lphasest}).
Although they (the black symbols) are initially not located in the regions that are rich in light nuclei,
the turbulence in the non-linear SASI broadens the distributions.
Figure~\ref{frahea52} shows the mass fractions and
 the heating rates of  different nuclear species  along the radial ray with $\theta =180^{\circ}$ at $t =$ 200~ms and another one with $\theta = 0^{\circ}$ at $t =$ 300~ms.
The heating rate of deuterons 
becomes as high as $\sim$ $10 \ \%$ of that of nucleons at $t=$ 200~ms around the bottom of the gain region. 
It should be noted that the cooling is subtracted for nucleons in $Q_E$,
whereas it is not included in $Q_{d,t,h,\alpha}$  because of the lack of the rates in the literature.
There are indeed large cancellations between heating and cooling at the bottom of the gain regions.
For instance, the heating and cooling rates per baryon for nucleons are 631 and $-$584 MeV/sec 
per baryon at $r=$ 110 km along the radial ray with $\theta =180^{\circ}$ at $t=$ 200~ms.
If we compare the pure heating rates, deuterons have 4.73 MeV/sec per baryon,
which is  just 0.75  $\%$ of the pure heating rate for nucleons  in the same example.
Around the same time the shock wave moves northwards ($\theta=0^{\circ}$) 
and the matter in the southern part $(\theta=180^{\circ})$  goes down deep into the central region.
The orange symbols in the top panel of Fig.~\ref{lphase52}  indicate that the matter in this southern part  has
low entropies, resulting in more deuterons  near the bottom of the gain region than the matter in other parts.
At $t=$ 300~ms, the shock wave reaches at $\sim$ 400 km 
and the heating of alpha particles is dominant for the same reasons we have explained in Fig.~\ref{fraheast}.
We can see in the bottom panel of Fig.~\ref{lphase52}  that the matter  along the radial ray with $\theta=0^{\circ}$  
has also lower entropies  and as a consequence deuterons and alpha particles are abundant
in the regions of high and low densities, respectively.
Both at $t=$ 200 and 300~ms, deuterons have the heating rates  comparable to those of  nucleons near the bottom of the gain regions.
The heating rates of alpha particles are $\sim$ 10 $\%$ of those of nucleons around the shock wave.

Figure \ref{timeev2d} displays the time evolutions of the integrated heating rate of each nuclear species
together with the shock and gain radii
along  the two radial rays with $\theta=0^{\circ}$ and $90^{\circ}$.
The integrated heating rate for  the specific direction is calculated
as $4 \pi \displaystyle \int_{r_g(\theta)}^{r_s(\theta)} Q_{i}(r,\theta) r^2\rm{d}\it{r}$ 
in the same way as  for the 1D model in Fig.~\ref{timeev1d}.
The heating rates of deuterons and alpha particles are about 1 $\%$ and   0.1 $\%$, respectively, of those of nucleons
 in the linear phase of SASI.
However, both of them are much more efficient in the subsequent non-linear evolutions of  SASI  and shock revival.
We can see that deuterons and alpha particles 
have different temporal variations in the heating rates, 
 which is ascribed to the fact that they occupy different parts of the gain region as explained in section \ref{sec1D}.
Note that the heating of light nuclei occurs quite inhomogeneously and the local heating can be more efficient
 than the average as shown in Fig.~\ref{frahea52}.

 In the models explored so far, the temperature of $\nu_{\mu}$ is assumed to be 10 MeV. 
However, this value may be too high. In fact, recent simulations tend to predict the  $\nu_{\mu}$ temperatures much closer to those of 
$\nu_{e}$ and  $\bar{\nu}_e$ (e.g. \citet{Janka12}). 
We hence repeat some simulations with  $T_{\nu_{\mu}}=$ 5 MeV. As shown in Tab. 1, the  $\nu_{\mu}$-heating rate per baryon is 
reduced by a factor of~10. The decrease in the net heating rates is particularly severe for tritons, helions and alpha particles,
 since NC interactions with  $\nu_{\mu}$ are dominant for the heating of these species.
On the other hand, the total heating rates of deuterons are reduced by only 24$\ \%$
 because CC interactions with $\nu_e$ and $\bar{\nu}_e$ are more important. 
Figure~\ref{radievot5} shows the temporal evolutions of the average shock radii for the models with $L_{52}=5.2$ and $\dot{M}_{sun}=1.0$, 
the counter parts of those presented in Fig.~\ref{radievo}. The results are qualitatively different. In fact, 
shock revival occurs earlier without the heating of deuterons although the shock radius is larger at $t \sim \ 200$~ms with the deuteron-heating.
 This occurs because the heating of nucleons is reduced in the presence of deuterons during the shock expansion at $t\sim200$~ms, 
which is similar to what we observed in the models for $T_{\nu_{\mu}}=$ 10 MeV with $L_{52}=6.0$ presented in Fig.~\ref{radievo}.
 The difference between the models with and without alpha particles is smaller for $T_{\nu_{\mu}}=$ 5 MeV than for $T_{\nu_{\mu}}=$ 10 MeV. 
 Figure~\ref{fraheat5}, the counter part of Fig.~\ref{frahea52}, displays
 the mass fractions and heating rates of different light nuclei along the radial ray with $\theta =180^{\circ}$ at $t =$ 200~ms. 
Although the dynamics is stochastic owing to the turbulence induced by SASI and
 the shock positions are different between the models with $T_{\nu_\mu}$ = 5 MeV and 10 MeV, 
 the heating by deuterons are not so significantly decreased thanks to the  CC contributions, 
whereas the heating rates of alpha particles are diminished by a factor of ~10.
Provided the high energy-dependence of the $\nu_{\mu}$-heating, we think that our standard models with $T_{\nu_{\mu}}=$ 10 MeV
 give the upper limit of the  $\nu_{\mu}$-heating whereas the models with  $T_{\nu_{\mu}}=$ 5 MeV will set the lower limit.

 In our standard models, the CC interactions on light nuclei are ignored in
 the temporal evolution of electron fraction. One may be worried, however,
 that this could affect the dynamics, since the CC interactions are dominant 
in the heating of deuterons. Note that the CC contributions are much smaller
 for $t, h$ and $\alpha$. In order to address this issue, we have included the contributions
 of deuterons and tritons in Eq.~(\ref{eq:yeevo}), the former of which is given by 
 \begin{eqnarray}
 Q_{Nd}  = X_{d}  \frac{31.6 \ {\rm s^{-1}}}{(r/10^{7} \rm{cm})^{2}}
  &&\left[
     \frac{L_{\nu_{\rm{e}}}}{10^{52}\rm{ergs~s}^{-1}}
     \left(\frac{5\rm{MeV}}{T_{\nu_{\rm{e}}}}\right)
     \frac{A_d^{-1}\langle
     \sigma_{\nu_{\rm{e}}}^{+}
     \rangle_{T_{\nu_{\rm{e}}}}}
     {10^{-40}\rm{cm}^2} \right. \nonumber \\
 &&-\left.\quad\frac{L_{\bar{\nu}_{\rm{e}}}}{10^{52}\rm{ergs~s}^{-1}}
  \left(\frac{5\rm{MeV}}{T_{\bar{\nu}_{\rm{e}}}}\right)
  \frac{A_d^{-1}\langle
  \sigma_{\bar{\nu}_{\rm{e}}}^{-}
   \rangle_{T_{\bar{\nu}_{\rm{e}}}}}
  {10^{-40}\rm{cm}^2} \right].
 \label{eq:light2}
\end{eqnarray} 
This expression is obtained by just replacing the energy-transfer coefficient 
$\langle \sigma_{\nu}^{\pm} E_{\nu} \rangle_{T_{\nu}}$, 
with the cross section averaged over the neutrino spectrum $\langle \sigma_{\nu}^{\pm}\rangle_{T_{\nu}}$, in Eq.~(\ref{eq:light}).
 We ignore the other contributions.
 For tritons, only the electron-type anti-neutrinos are included, 
since the cross section for $\nu_e$ is currently unavailable.
 Note, however, that the contribution of tritons is negligible anyway as shown shortly. 

 We initially construct the spherically symmetric, 
steady accretion flows and then perform 2D simulations with 
$Q_{Nd}$ and $Q_{Nt}$ being incorporated.
The neutrino luminosity and mass accretion rate are fixed to $L_{52}=5.2$ and $\dot{M}_{sun}=1.0$
 here. 
It turns out that the temporal evolution of the shock wave is unchanged from that presented in Fig. 6, 
which justifies the neglect of these effects in the standard models. 
In fact, Fig.~\ref{CCcheck} 
demonstrates by comparing $Q_{Nd}$ and $Q_{Nt}$  with $Q_N$,
 the contribution of nucleons, at $t=$ 0, 200~ms that
 $Q_{Nd}$ is about 0.1 $\%$ of $Q_N$
 in most regions and $Q_{Nt}$ is always negligibly small.
Although we can see in the initial condition a slight increase in $Y_e$ by the inclusion of $Q_{Nd}$ and $Q_{Nt}$,
 the difference disappears by $t = 200$~ms. 

 The reasons why deuterons are non-negligible heating sources, giving 1-10 $\%$
 of the nucleon contributions, but they do not contribute to the evolution of $Y_e$
 are the following: (1) the net heating rate of nucleons, $Q_E$, is an outcome of rather big cancellations between heating and cooling
 near the gain radius whereas such cancellations are not so large in $Q_N$; as a result $Q_N$ is much greater than $Q_{Nd}$
 in the region where the deuteron-heating is efficient;
 (2) the $Y_e$ evolution is controlled by the competition between the electron-type neutrinos and anti-neutrinos; 
according to the current formula, $Q_{Nd}$ is given as
 $Q_{Nd} =Q_{Nd}(\nu_e d)- Q_{Nd}(\bar{\nu}_e d) = 0.188 \ Q_{Nd}(\nu_e d)$
 and the cancellation is much larger than for $Q_N$;
(3) the NC reactions contribute 12 and 33 $\%$ to $Q_d$ for $T_{\nu_\mu}=5$ and 10 MeV, respectively, but none to $Q_{Nd}$.

To summarize, the heating via light nuclei is non-negligible in the non-linear phase of SASI, in which various values of $(\rho, T)$ are realized,
 whereas they are minor in the linear stage.
Among light nuclei, deuterons play important roles near the bottom of the gain region 
whereas alpha particles are influential near the shock fronts when the shock wave is expanding and, as a consequence, 
the densities and temperatures become lower.
 The results are rather sensitive to the $\nu_\mu$ spectrum. 
If the temperature of $\nu_{\mu}$ is as low as that of $\bar{\nu}_e$, 
the heating of alpha particles will be substantially diminished whereas the deuteron-heating is not so much reduced.

\section{Summary and Discussions}
We have investigated the influences of the inelastic interactions of neutrinos with light nuclei on the dynamics in the  post-bounce phase of core-collapse supernovae.
We have done 2D numerical simulations of SASI with the assumption of axial symmetry for some representative combinations of the luminosity and mass accretion rate.
We have not solved the dynamics of the central part of the core and replaced it with the suitable boundary conditions
 and have started the simulations from spherically symmetric steady state, adding some perturbations to the radial velocity.
The neutrino transport has been handled by the simple light-bulb approximation with the time-independent Fermi-Dirac spectrum. 
In addition to the ordinary heating and cooling reactions with nucleons,
 we have taken into account the heating reactions with four light nuclei for the first time.
The abundance of light nuclei is provided by the multi-nuclei EOS together with other thermodynamical quantities.

 We have found that 
the evolutions of shock waves in 2D are influenced by the heating of deuterons and alpha particles
and that they have different roles.
In the initial steady states, the heating by light nuclei is the most efficient
for the combination of high neutrino luminosity and low mass accretion rate,
since the shock radius is large and the matter near the shock front has the low densities
 and temperatures that yield a large amount of alpha particles.
On the other hand, deuterons are populated near the bottom of the gain region, where matter has higher densities and temperatures.
They hence have some impacts on the shock radius regardless of neutrino luminosities and mass accretion rates.
From the results of 1D simulations, we have found
 that the integrated heating rates of deuterons and alpha particles become high at different phases
 in the oscillations of shock wave: 
 the heating rate of deuterons becomes the highest when the shock radius hits the minimum and
 the matter compression is the greatest.
Whereas the heating via deuterons is constantly effective,
the heating of alpha particles becomes important 
only when the shock wave has large radii and matter has low entropies.
%
The dynamics in 2D is more sensitive to the inclusion of the light-nuclei heating
because SASI in the non-linear regime makes the gain region more inhomogeneous and 
there appear the regions that have densities and temperatures favorable for the existence of light nuclei.
The heating rates of light nuclei reach about 10 $\%$ of those of nucleons locally.
 As a consequence, the dynamics of shock revival is influenced by the heating via light nuclei.
 In particular, the heating by deuterons brings non-negligible changes, which may be positive or negative for shock revival,
 when the neutrino luminosity is close to the critical value.
 The results are rather sensitive to the neutrino spectrum.
 In the case of  $T_{\nu_{\mu}}=5$ MeV
 instead of  $T_{\nu_{\mu}}=$10 MeV, 
the heating of alpha particles is reduced by $\sim 90 \ \%$ 
whereas the heating via deuterons are not so much affected,
 since the CC reactions with $\nu_e$ and $\bar{\nu}_e$ are more important for the deuteron-heating.
%

\begin{figure}
\begin{center}   
\begin{tabular}{ll}
\epsscale{1.100}
   \plottwo{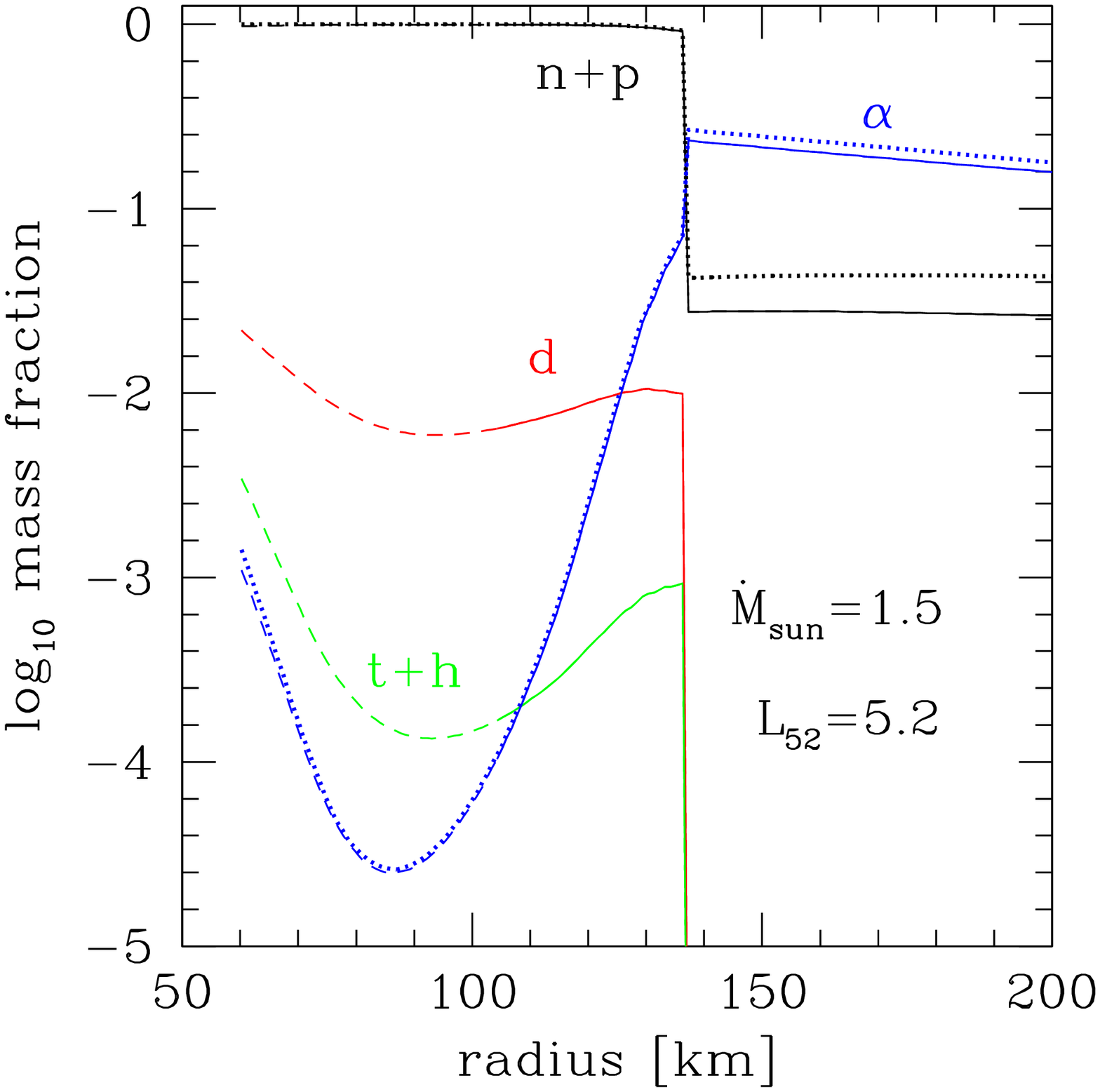}{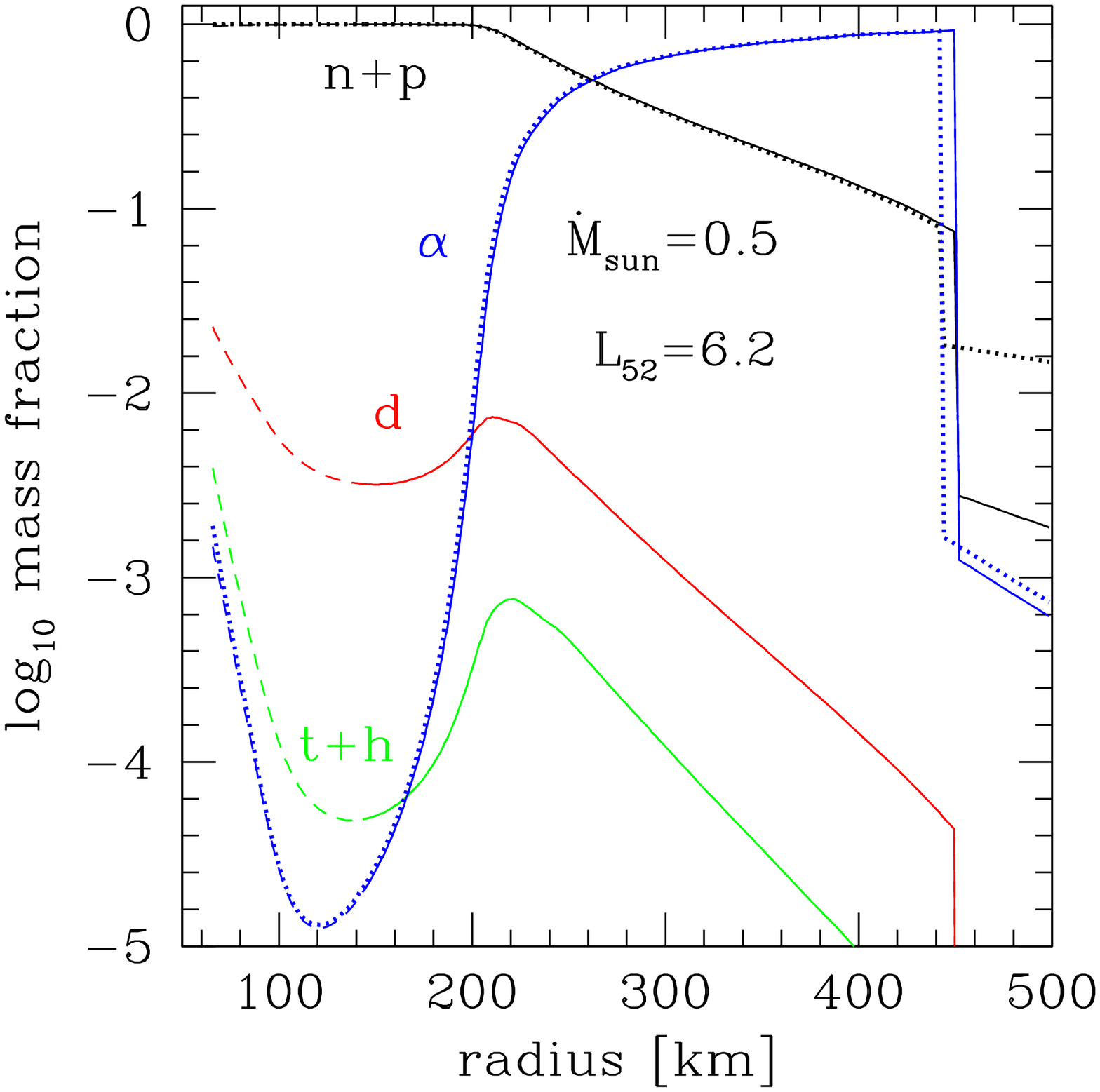} \\
   \plottwo{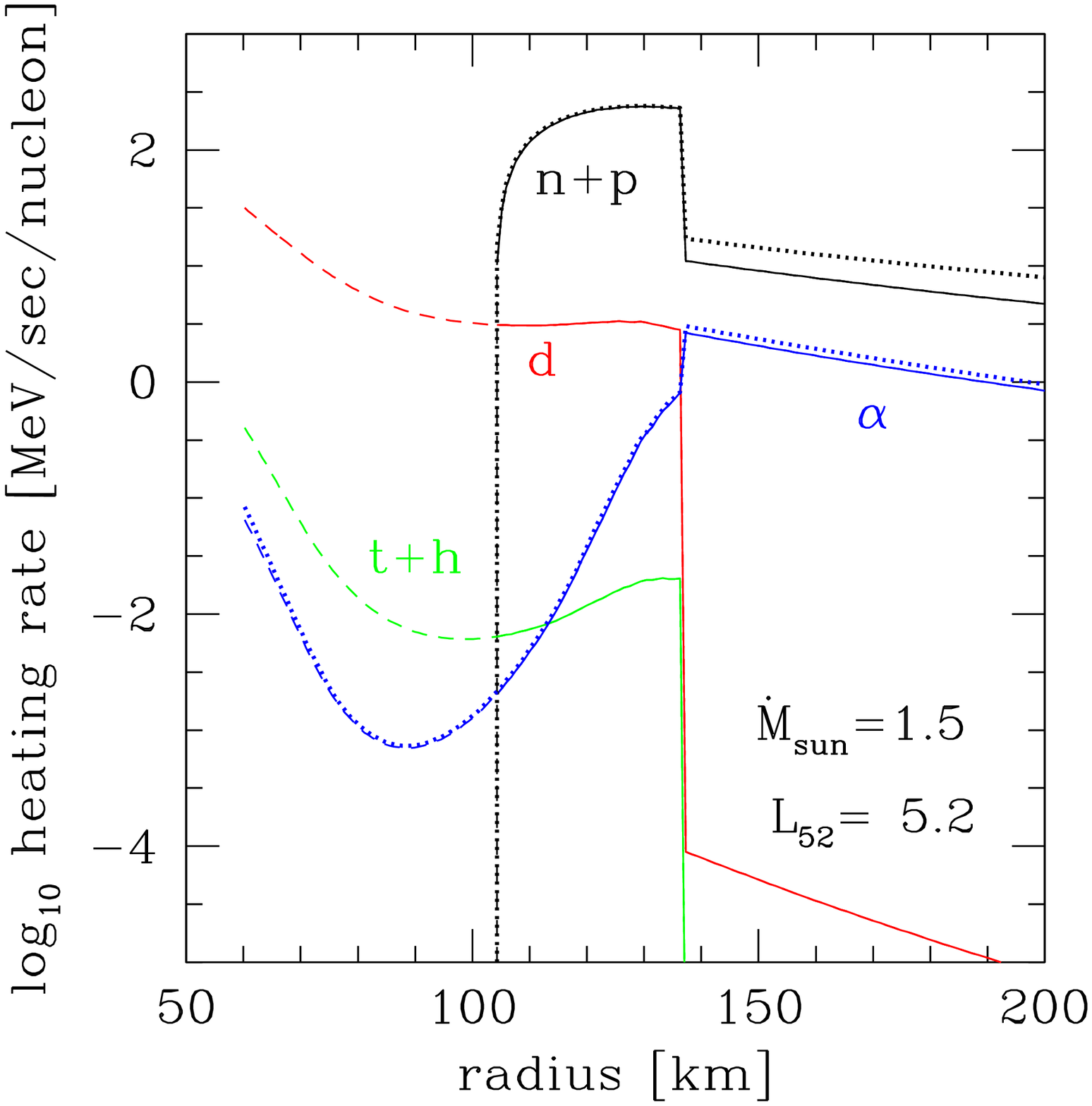}{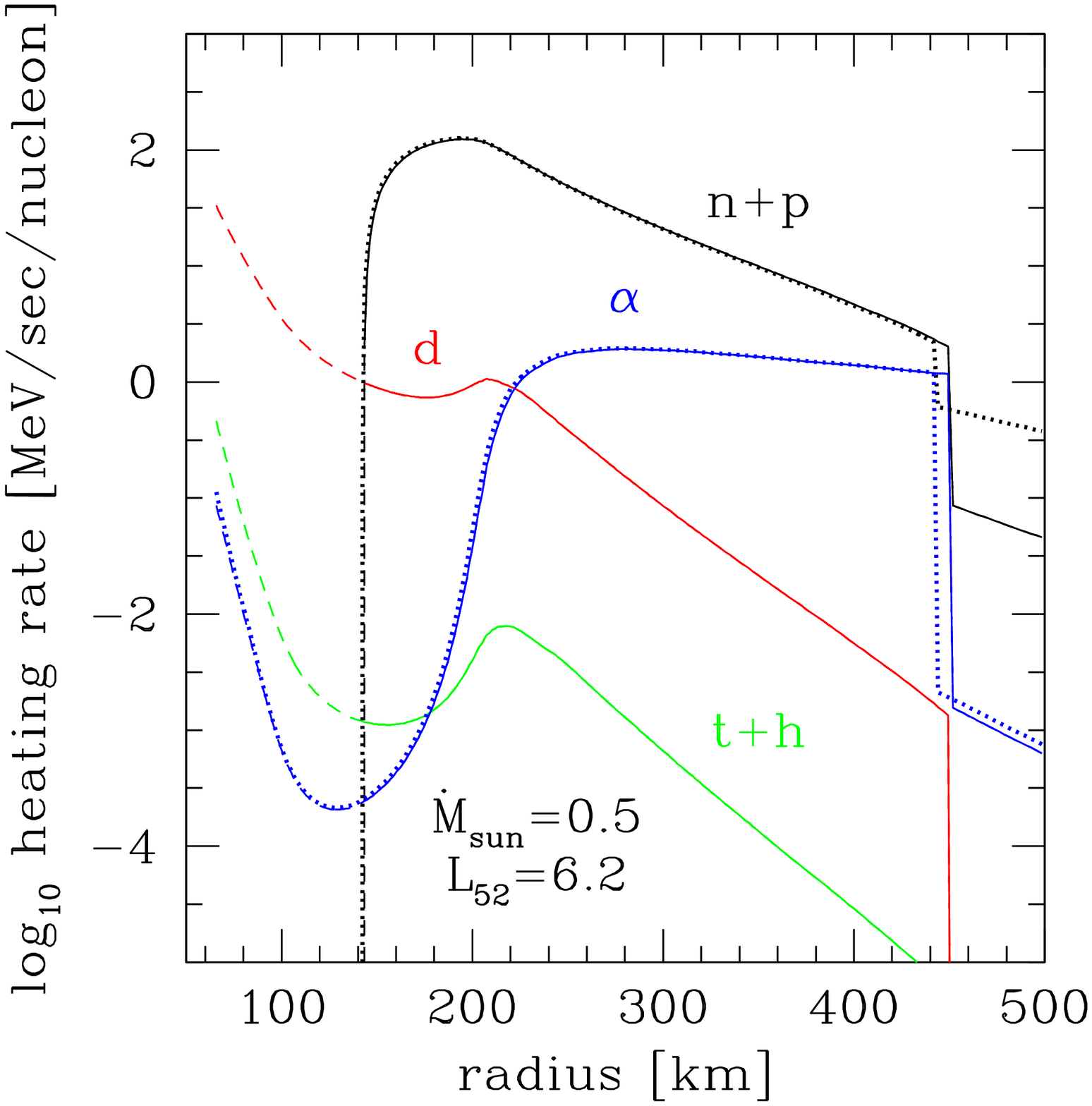} \\
    \end{tabular}
\end{center}
\caption{The mass fractions (upper) and heating rates per baryon (bottom) of
$A_i= 1$: protons and neutrons (black), $A_i= 2$: deuterons (red), $A_i= 3$: tritons and helions (green) and $A_i=4$: alpha particles (blue)
for $L_{52}=5.2$ with $\dot{M}_{sun}=1.5$   (left panel)  and   $L_{52}=6.2$  with $\dot{M}_{sun}=0.5$   (right panel).
Dashed lines indicate the cooling regions where the cooling reaction of nucleons is dominant.
 Dotted lines are the results of  the models with the Shen's EOS for the same $L_{52}$ and $\dot{M}_{sun}$.}
\label{fraheast}
\end{figure}

\begin{figure}
\begin{center}   
 \begin{tabular}{c}
               \resizebox{120mm}{!}{\plotone{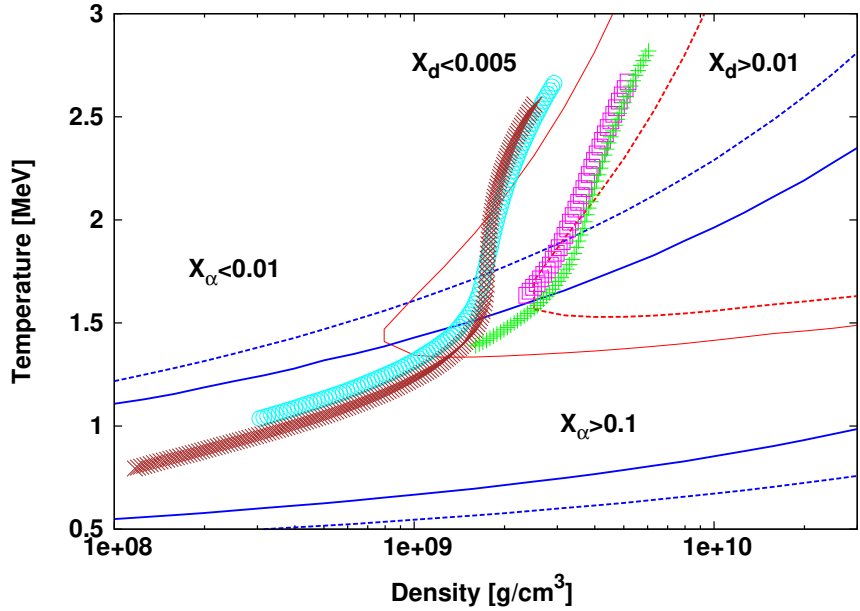}} \\
    \end{tabular}	
\end{center}
\caption{The lines show the  contours of mass fractions of deuteron $X_d$ and alpha particle $X_{\alpha}$   at $Y_e=0.5$ (red thin lines for $X_d=0.005$, 
red dashed lines for $X_d=0.01$, blue dashed lines for $X_{\alpha}=0.01$ and blue solid lines for $X_{\alpha}=0.1$).
The symbols mean the densities and temperatures in the gain regions for 
$L_{52}=5.2$ with $\dot{M}_{sun}=0.5$  (cyan)  and 1.5  (magenta)  and $L_{52}=6.2$ with $\dot{M}_{sun}=0.5$  (brown)  and 1.5  (green).}
\label{lphasest}
\end{figure}

\begin{figure}
\begin{center}   
\begin{tabular}{c}
  \resizebox{110mm}{!}{\plotone{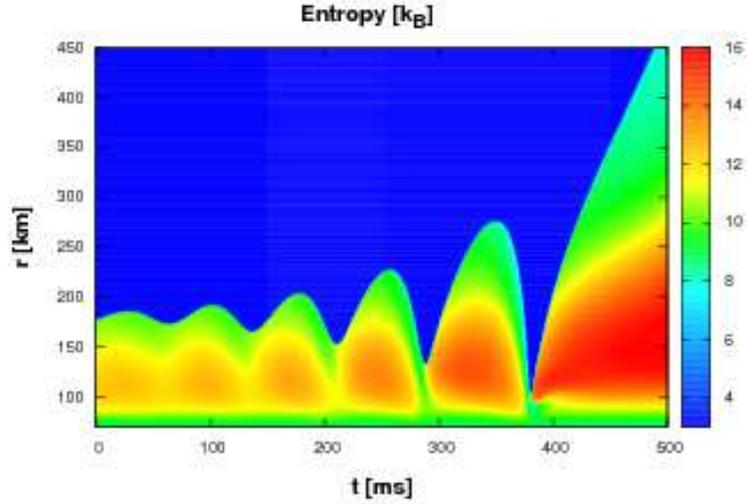}}  
    \end{tabular}
\end{center}
\caption{The entropy per baryon in  the $(t,r)$ plane for the 1D model with
$L_{52}=5.4$ and $\dot{M}_{sun}=$ 1.0.}
\label{timeen}
\end{figure}

\begin{figure}
\begin{center}   
\begin{tabular}{c}
  \resizebox{110mm}{!}{\plotone{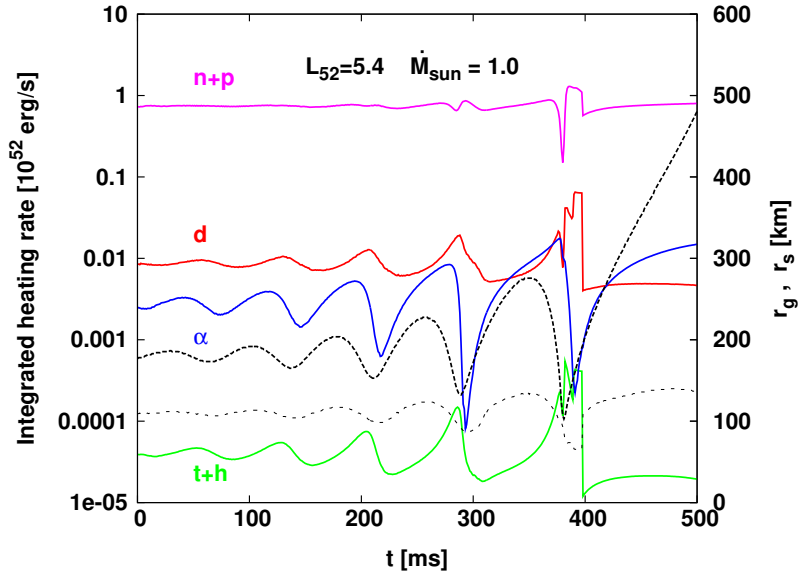}}  
    \end{tabular}
\end{center}
\caption{The time evolutions of the average shock and gain radii and
 integrated heating rates of different nuclear species.
 Black dashed and dotted lines denote the shock  and gain radii, respectively. 
 Magenta, red, green and  blue lines represent the heating rates of
 $A_i=1$ (nucleons), $A_i=2$ (deuterons), $A_i=3$ (tritons and helions)
 and $A_i=4$ (alpha particles), respectively.}
\label{timeev1d}
\end{figure}

\begin{figure}
\begin{center}   
\begin{tabular}{l}
 \resizebox{85mm}{!}{\plotone{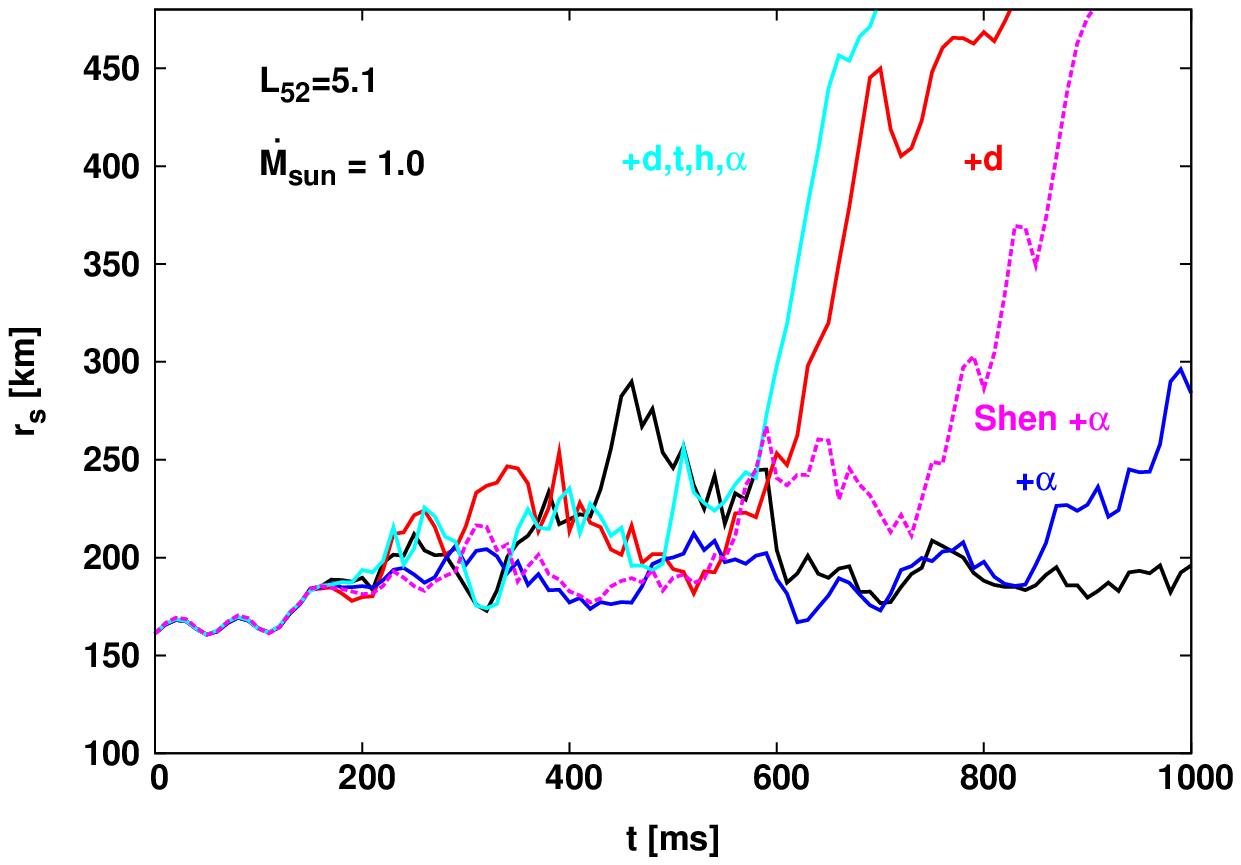}} 
 \resizebox{85mm}{!}{\plotone{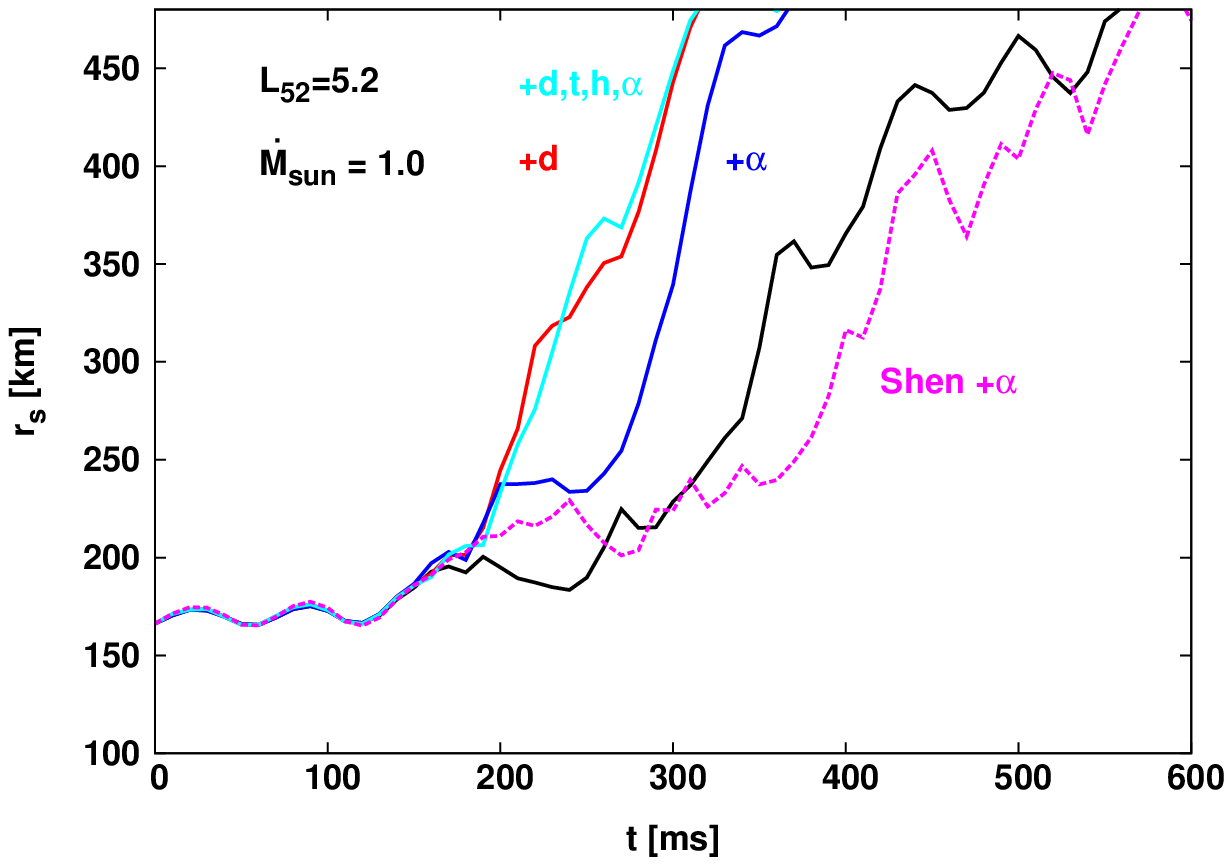}}  \\
 \resizebox{85mm}{!}{\plotone{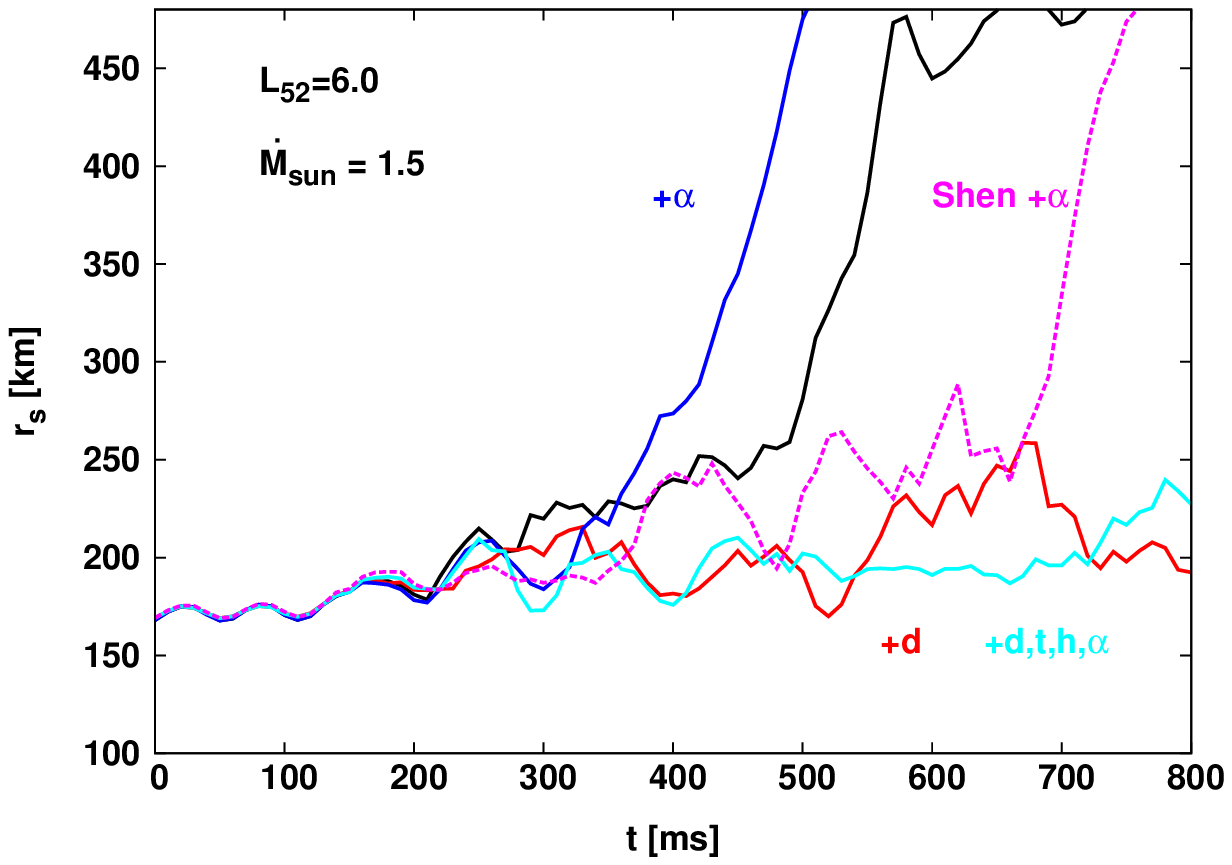}}  
 \resizebox{85mm}{!}{\plotone{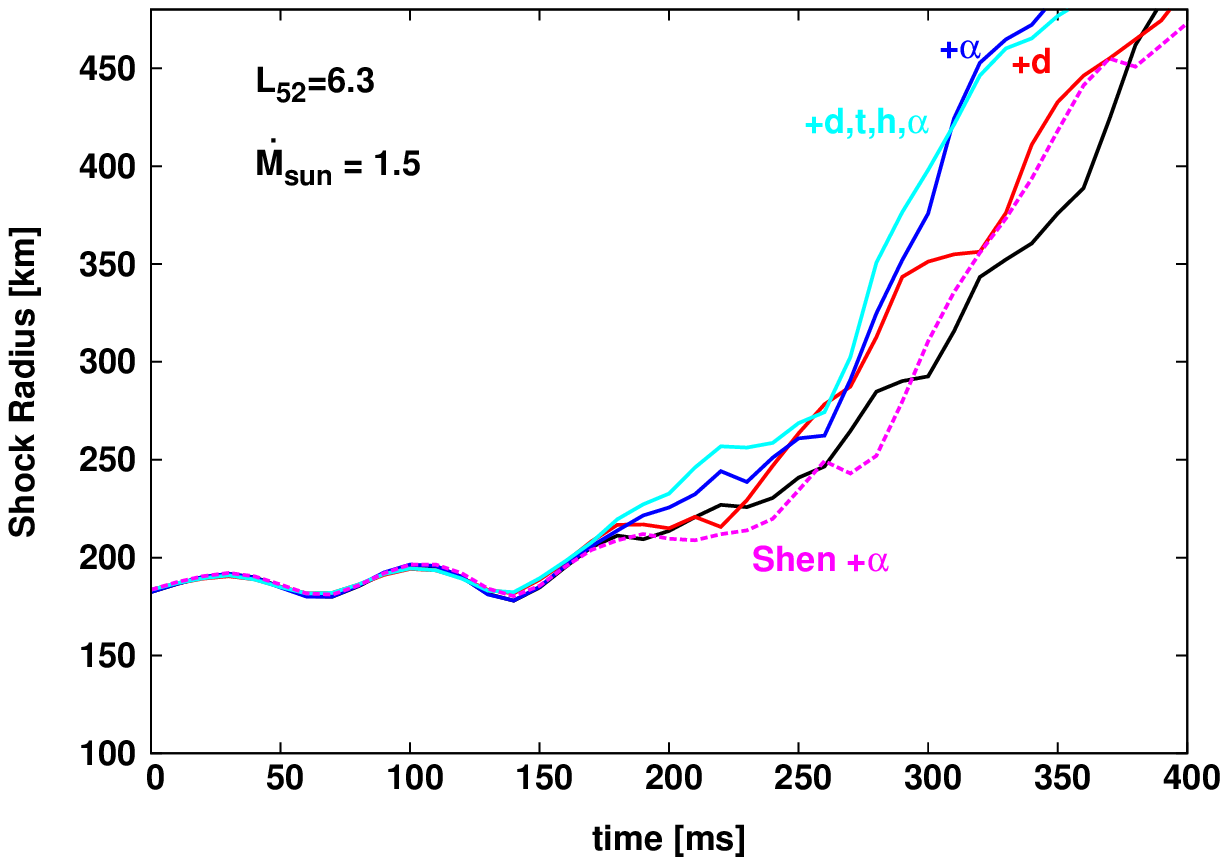}}  
    \end{tabular}
\end{center}
\caption{ The time evolutions of the average shock radii for the models with the heating of all light nuclei (cyan solid lines),
 only deuterons (red solid lines), only alpha particles (blue solid lines) and no light nuclei (black solid lines)
 as well as the models with the Shen's EOS and the heating of alpha particles (magenta dashed lines).
The combinations of the luminosity and mass accretion rate are $L_{52}=5.1$ and 5.2 with $\dot{M}_{sun}=1.0$ and 
$L_{52}=6.0$ and 6.3 with $\dot{M}_{sun}=1.5$.}
\label{radievo}
\end{figure}

\begin{figure}
\begin{center}   
\begin{tabular}{l}
  \resizebox{140mm}{!}{\plotone{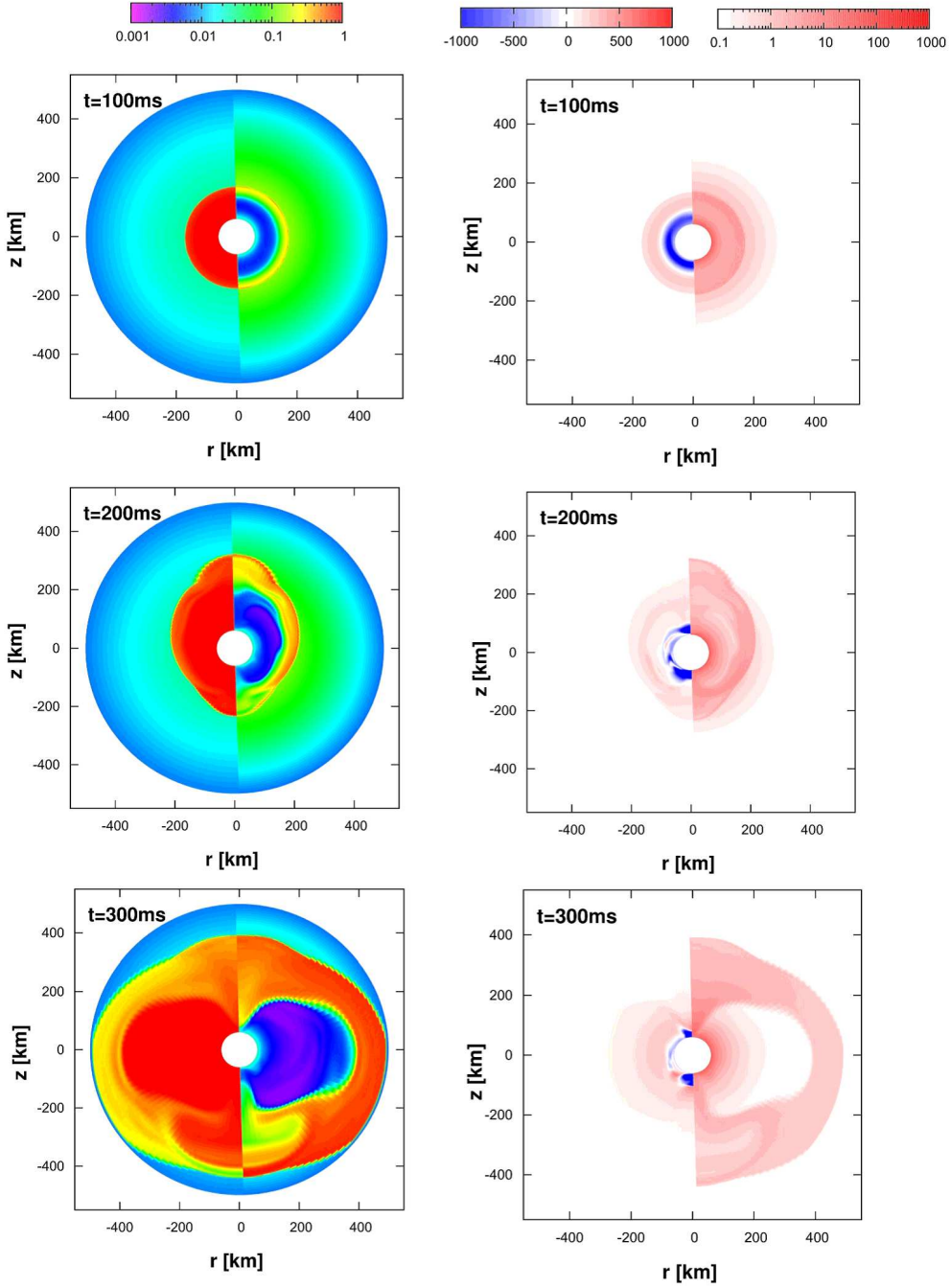}}  
\end{tabular}
\end{center}
\caption{The mass fractions of nucleons, $X_n+X_p$ (left halves of left panels), 
those of all light nuclei, $X_d+X_t+X_h+X_{\alpha}$ (right halves of left panels),
the heating and cooling  rates per baryon in the unit of MeV/sec of nucleons,
 $Q_E\times m_u/\rho$ (left halves of right panels), and those of  all light nuclei,
 $(Q_d+Q_t+Q_h+Q_{\alpha}) \times m_u/\rho$ (right halves of right panels).
The times are $t=$ 100, 200 and 300~ms and the luminosity and accretion rate are $L_{52}=5.2$ and $\dot{M}_{sun}=1.0$.}
\label{snap52}
\end{figure}

\begin{figure}
\begin{center}   
 \begin{tabular}{c}
                \resizebox{82mm}{!}{\plotone{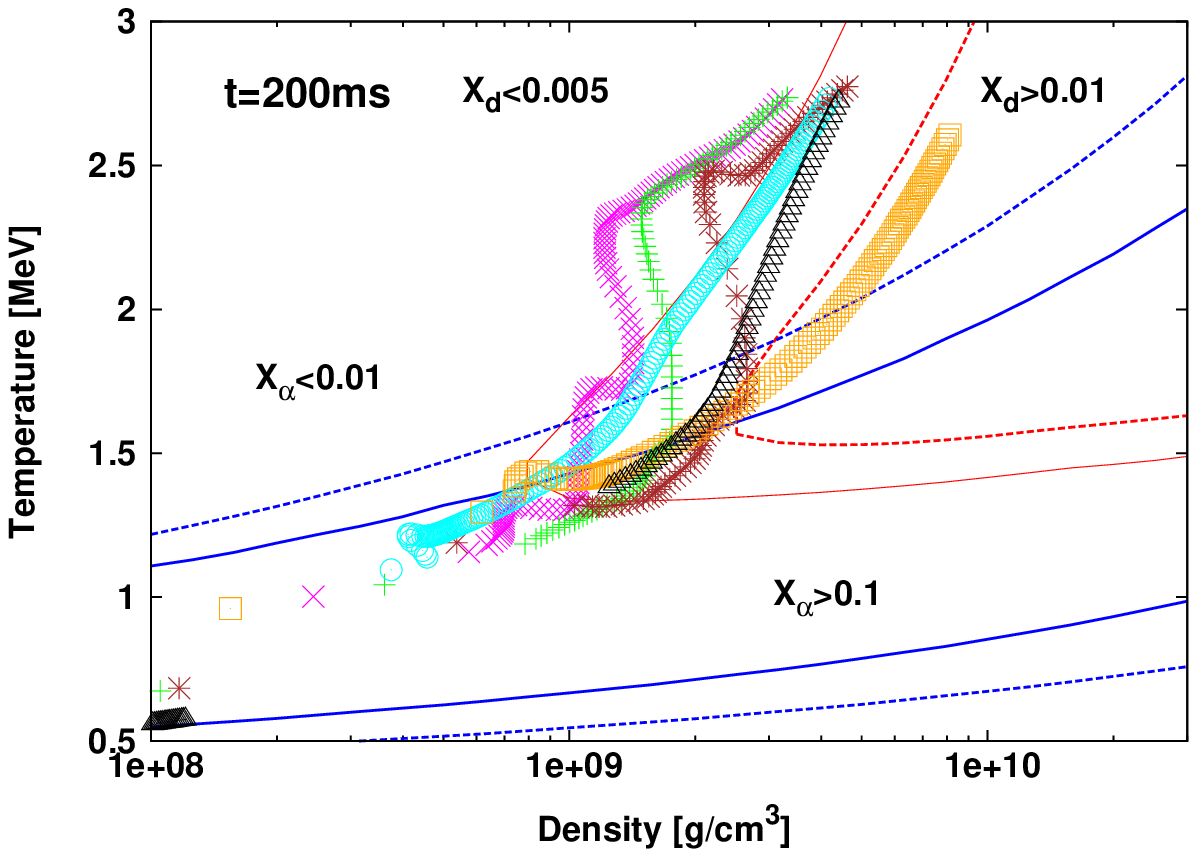}} \\
				\resizebox{82mm}{!}{\plotone{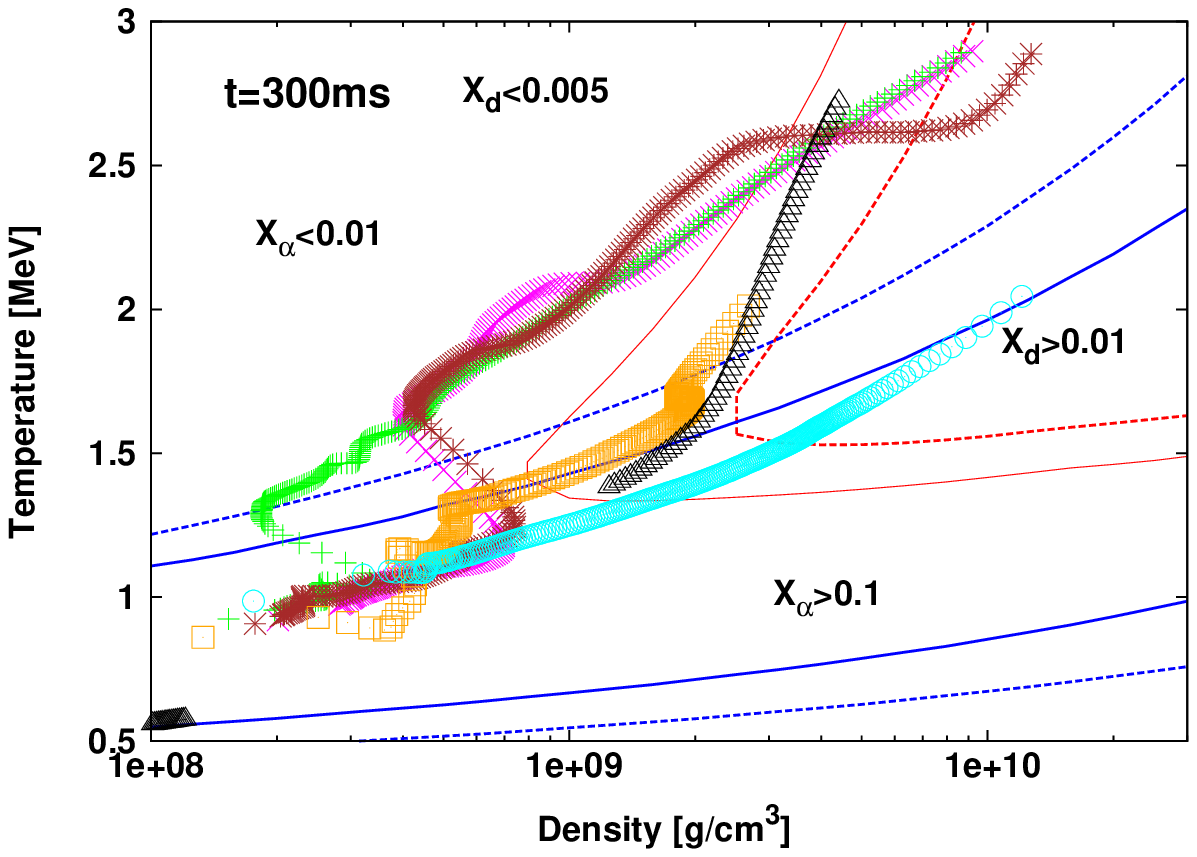}}  
    \end{tabular}	
\end{center}
\caption{The lines  are the same as in Fig.~\ref{lphasest}.
The symbols show the densities and temperatures in the gain regions at $t=$ 200~ms (upper panel) and 300~ms (lower panel) 
for the radial rays with
$\theta=0^{\circ}$ (cyan), $45^{\circ}$ (magenta), $90^{\circ}$  (green), $135^{\circ}$ (brown)  and $180^{\circ}$ (orange).
The black symbols correspond to the initial spherically symmetric state.}
\label{lphase52}
\end{figure}

\begin{figure}
\begin{center}   
\begin{tabular}{ll}
\epsscale{0.900}
   \plottwo{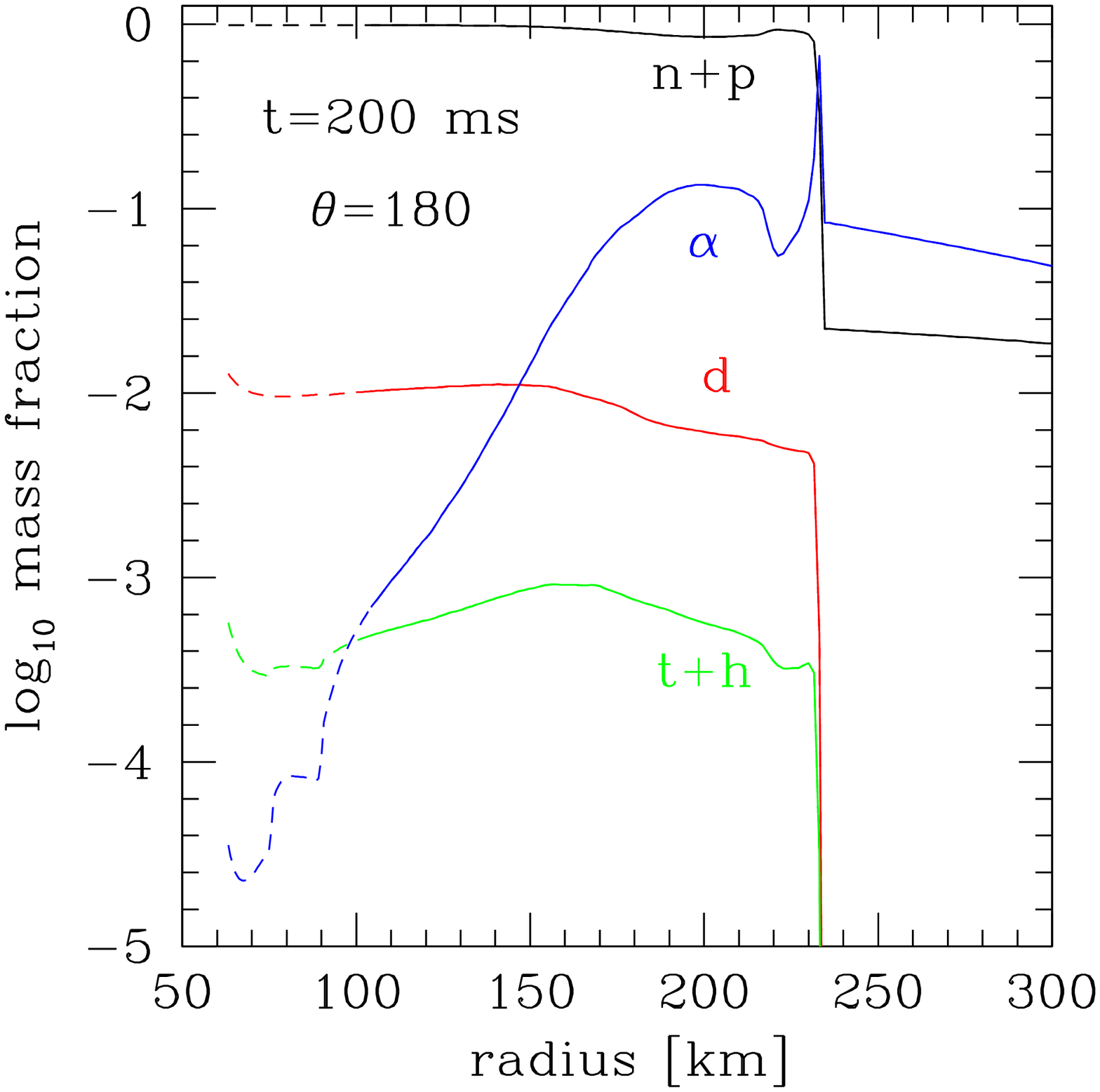}{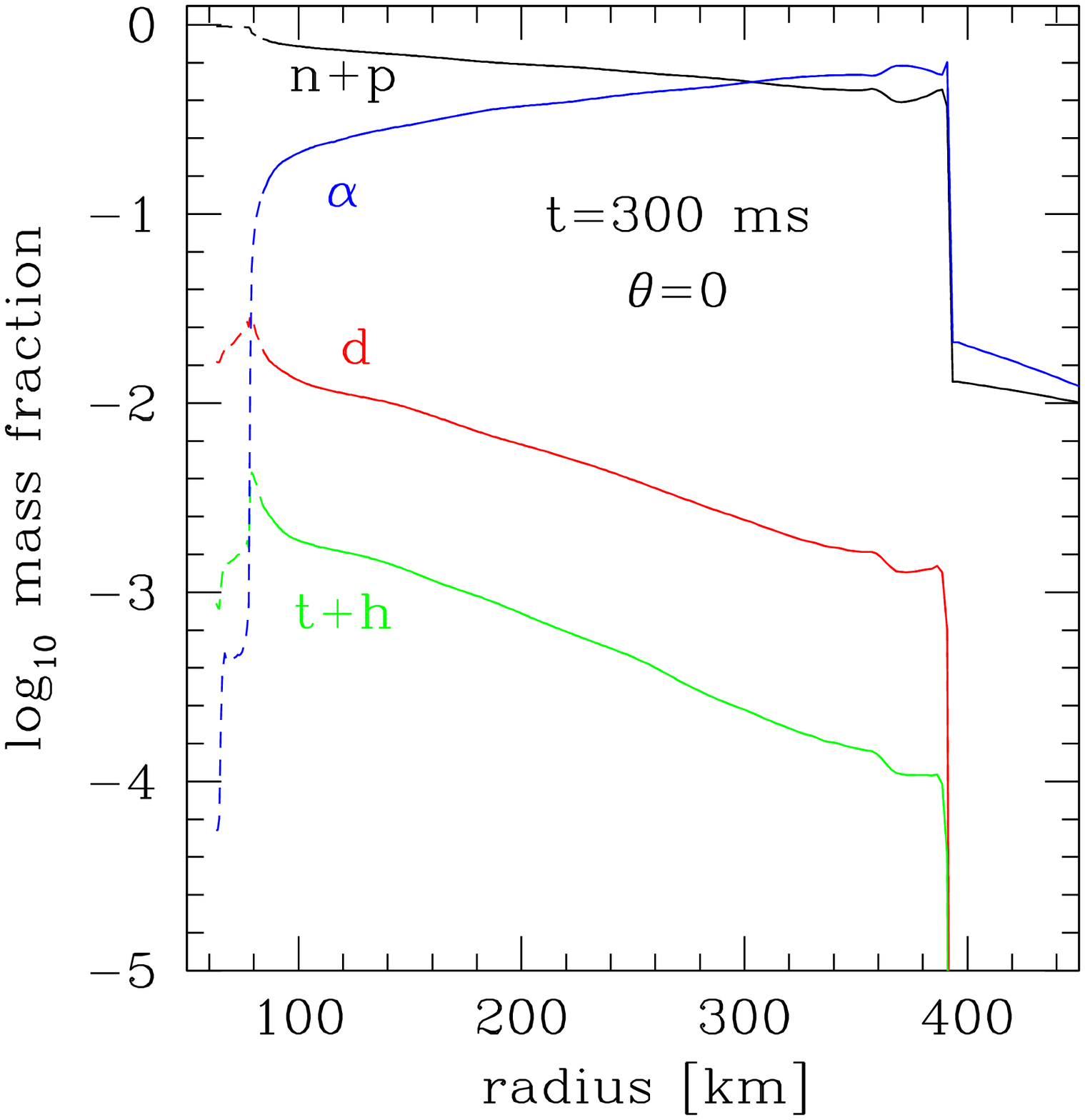} \\
   \plottwo{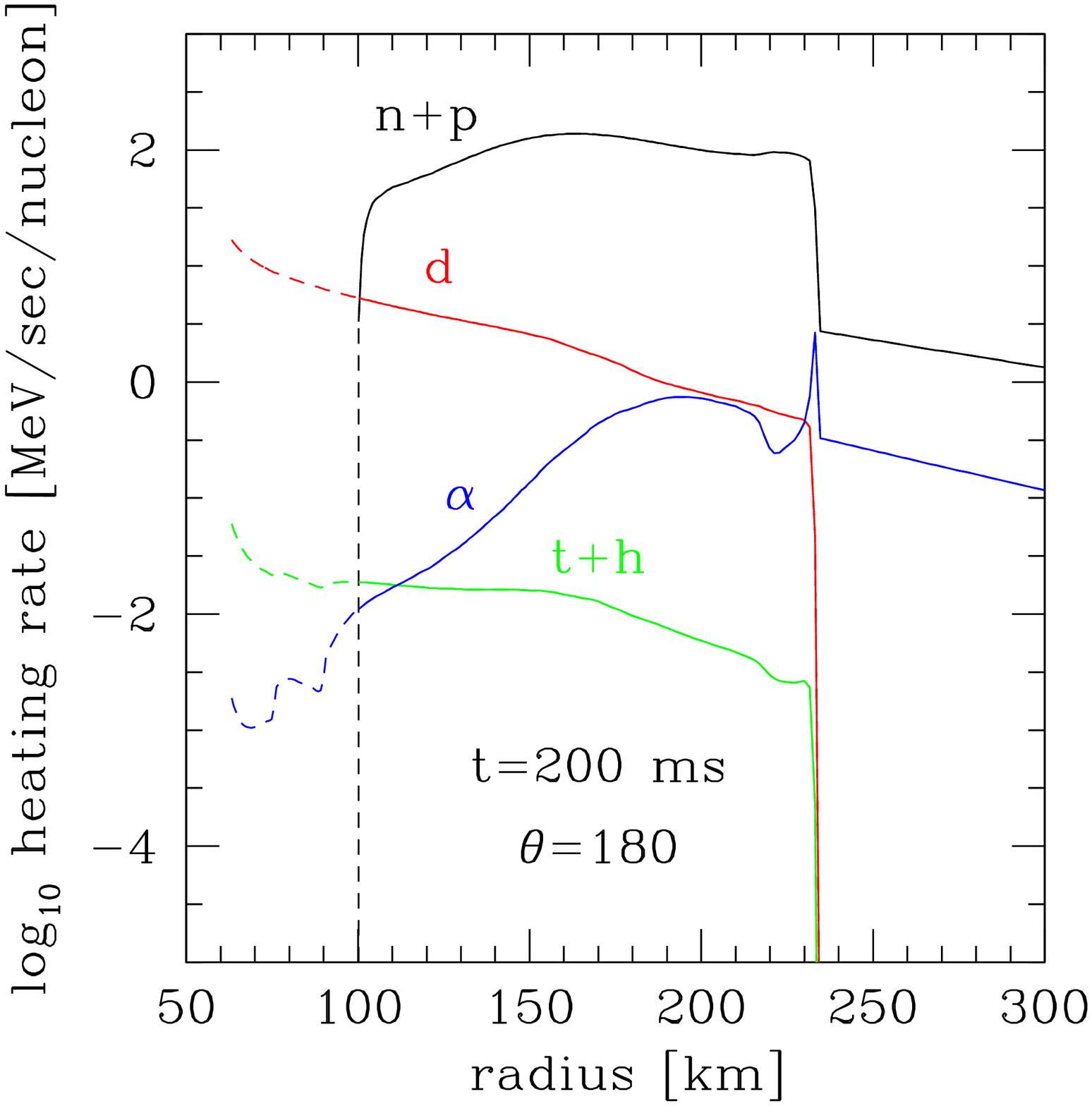}{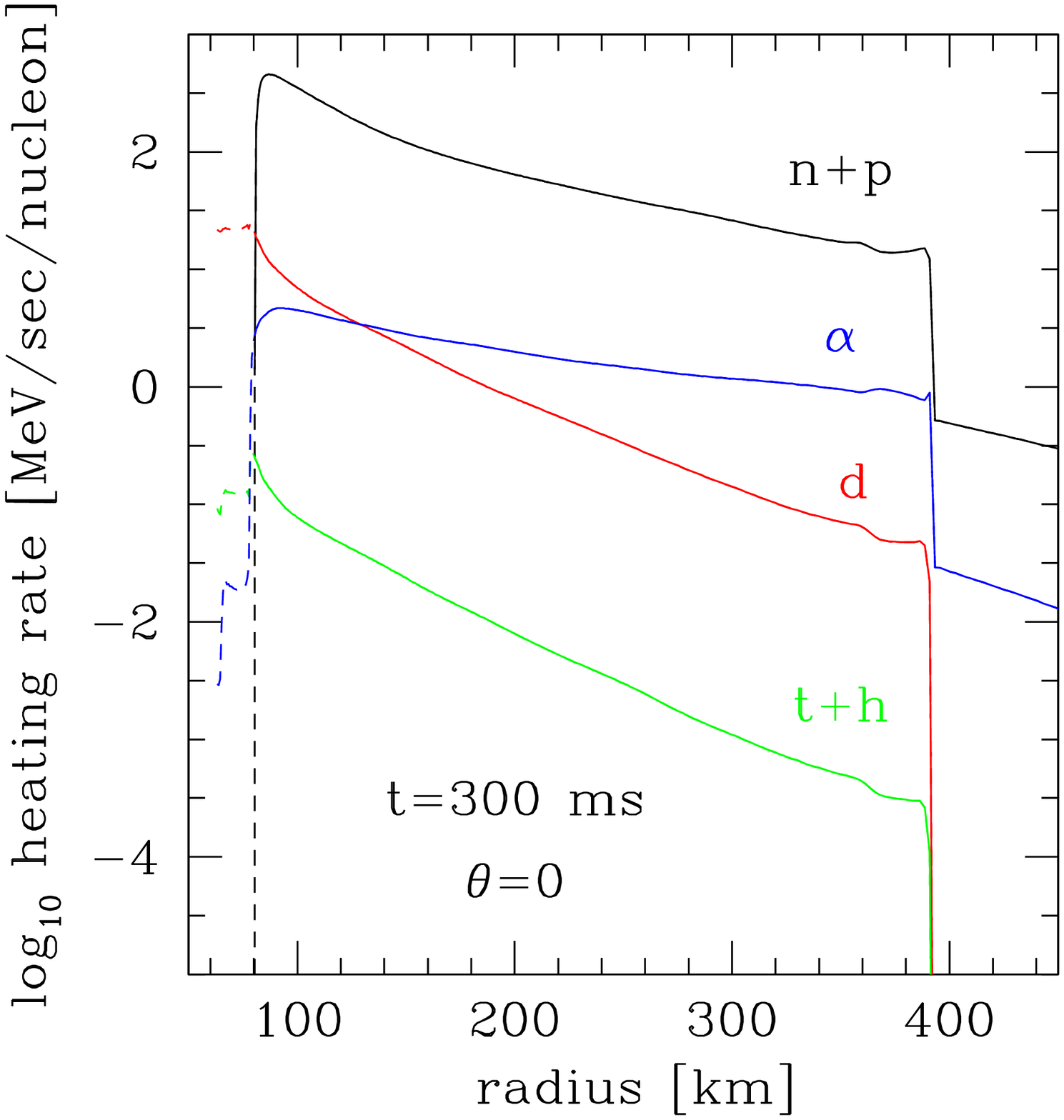} \\
    \end{tabular}
\end{center}
\caption{The mass fractions (upper panels) and heating rates per baryon (lower panels)
along the radial rays with $\theta=180^{\circ}$  at $t=200$ ms (left panels) 
and with $\theta=0^{\circ}$ at $t=300$~ms (right panels).
The notations of various lines are the same as in Fig.~\ref{fraheast}.}
\label{frahea52}
\end{figure}

\begin{figure}
\begin{center}   
\begin{tabular}{c}
  \resizebox{140mm}{!}{\plotone{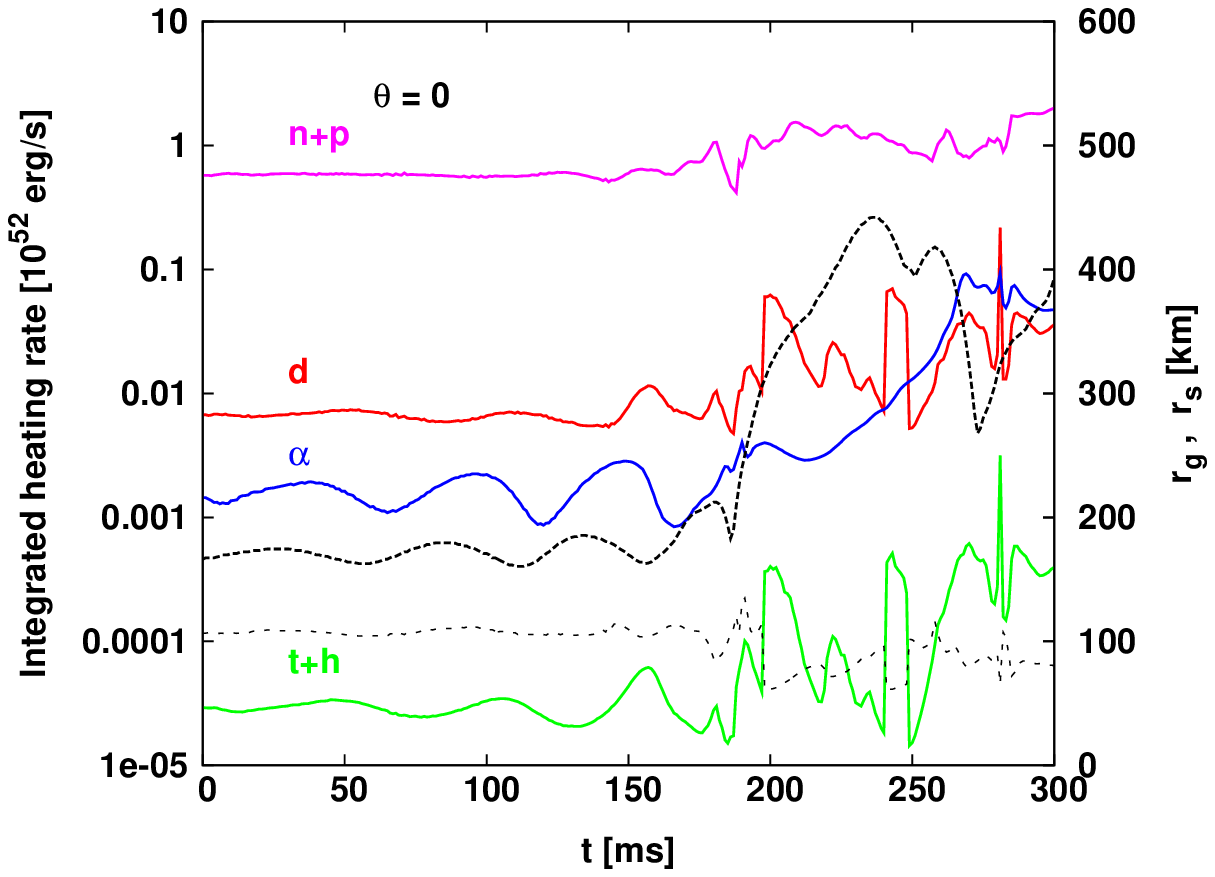}}  \\
  \resizebox{140mm}{!}{\plotone{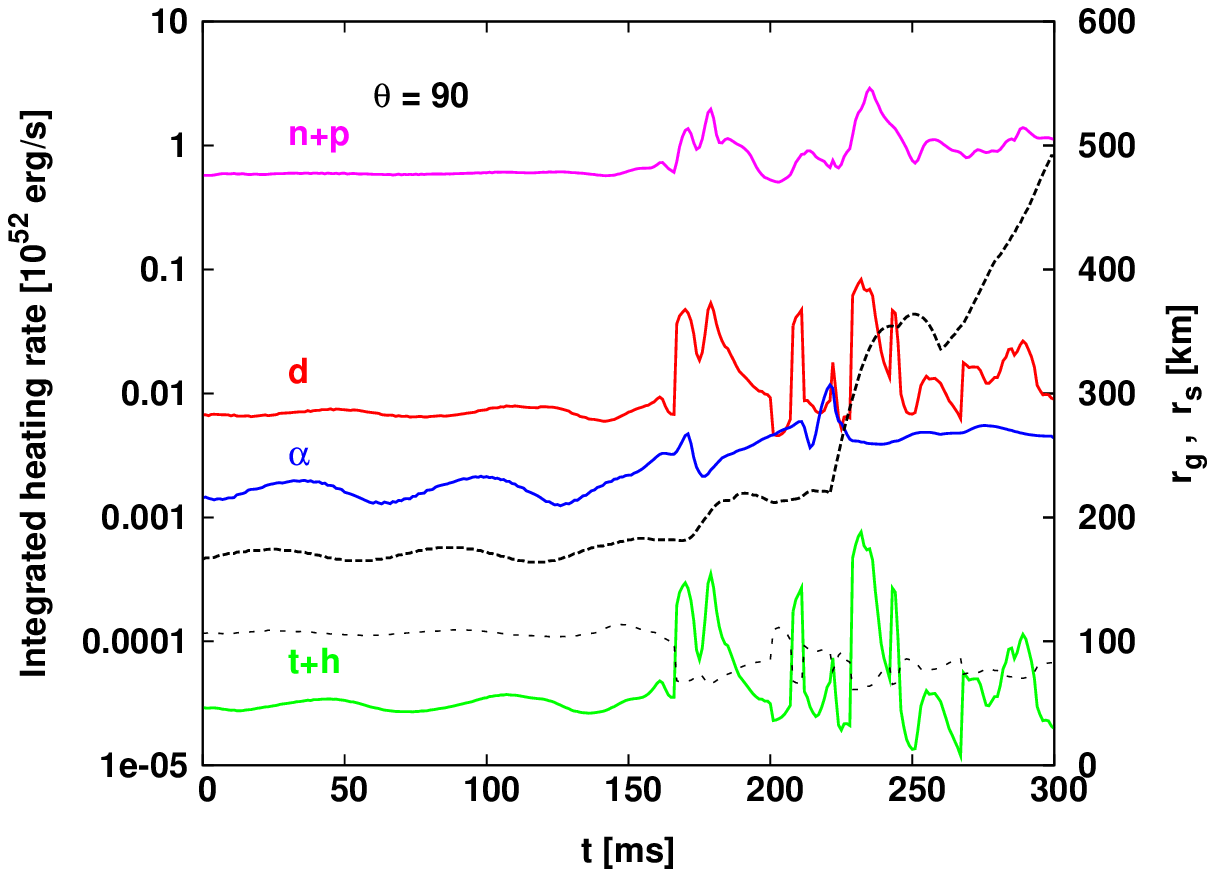}}  
    \end{tabular}
\end{center}
\caption{The time evolutions of the  shock and gain radii and integrated heating rates
on the radial rays with $\theta=0^{\circ}$ (upper panel) and $\theta=90^{\circ}$ (lower panel).
The notations of various lines are the same as in Fig.~\ref{timeev1d}.}
\label{timeev2d}
\end{figure}

\begin{figure}
\begin{center}   
\begin{tabular}{c}
 \resizebox{120mm}{!}{\plotone{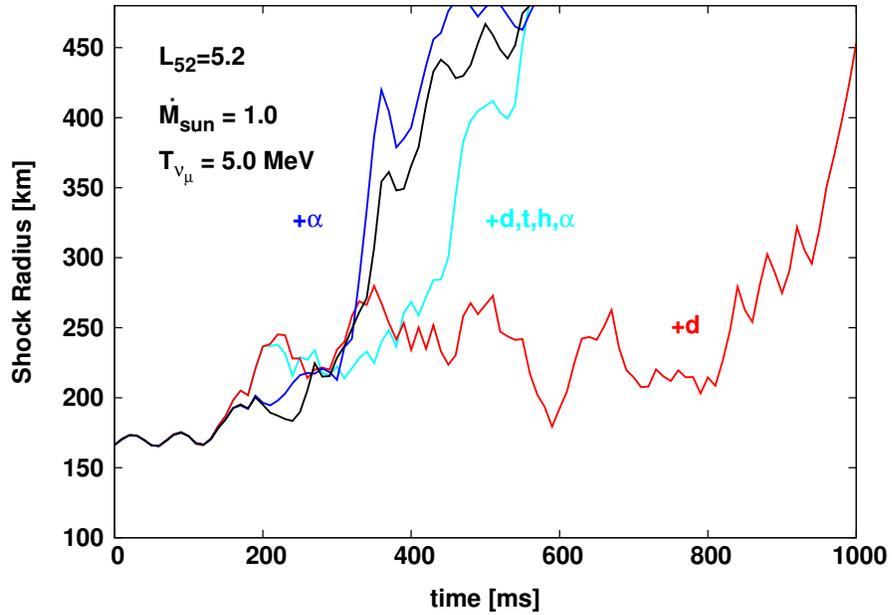}}  \\
    \end{tabular}
\end{center}
\caption{ The time evolutions of the average shock radii for the models with the heating of all light nucleus (cyan line),
 only deuterons (red line), only alpha particles (blue line) and no light nuclei (black line).
The temperature of $\nu_\mu$ is  $T_{\nu_{\mu}}=$ 5 MeV and the luminosity and mass accretion rate are 
$L_{52}=$ 5.2 and $\dot{M}_{sun}$=1.0, respectively.}
\label{radievot5}
\end{figure}

\begin{figure}
\begin{center}   
\begin{tabular}{l}
\epsscale{0.990}
   \plottwo{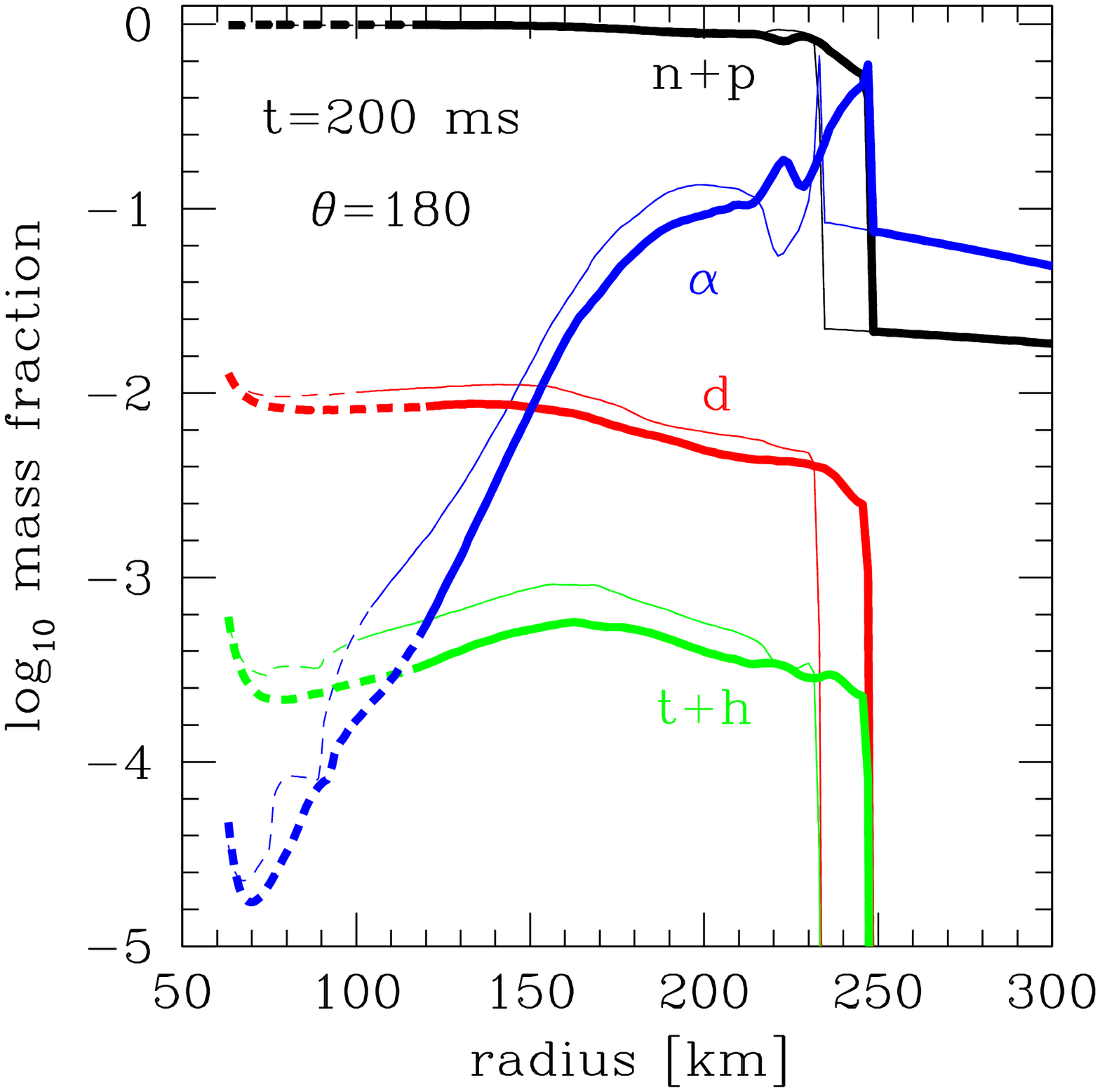}{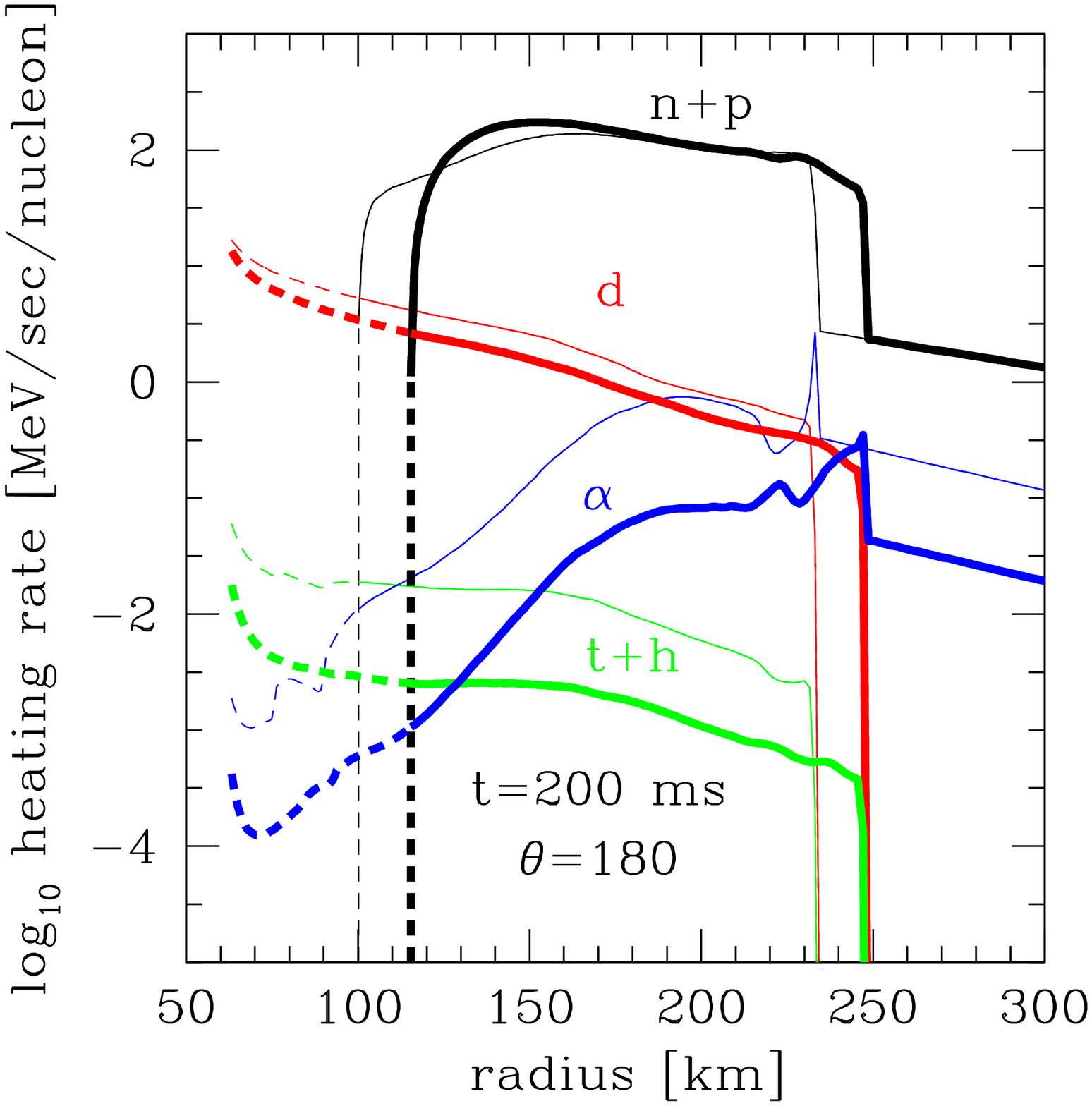}
    \end{tabular}
\end{center}
\caption{  The mass fractions (left panel) and heating rates per baryon (right panel)
along the radial ray with $\theta=180^{\circ}$ at $t=200$~ms for $T_{\nu_{\mu}}=5$ MeV (thick lines) and 10 MeV (thin lines).
The thin lines are just the same as those in the left panels of Fig~\ref{frahea52}
and the notations of lines are also identical to those  in Fig. \ref{frahea52}.}
\label{fraheat5}
\end{figure}

\begin{figure}
\begin{center}   
\begin{tabular}{c}
  \resizebox{140mm}{!}{\plotone{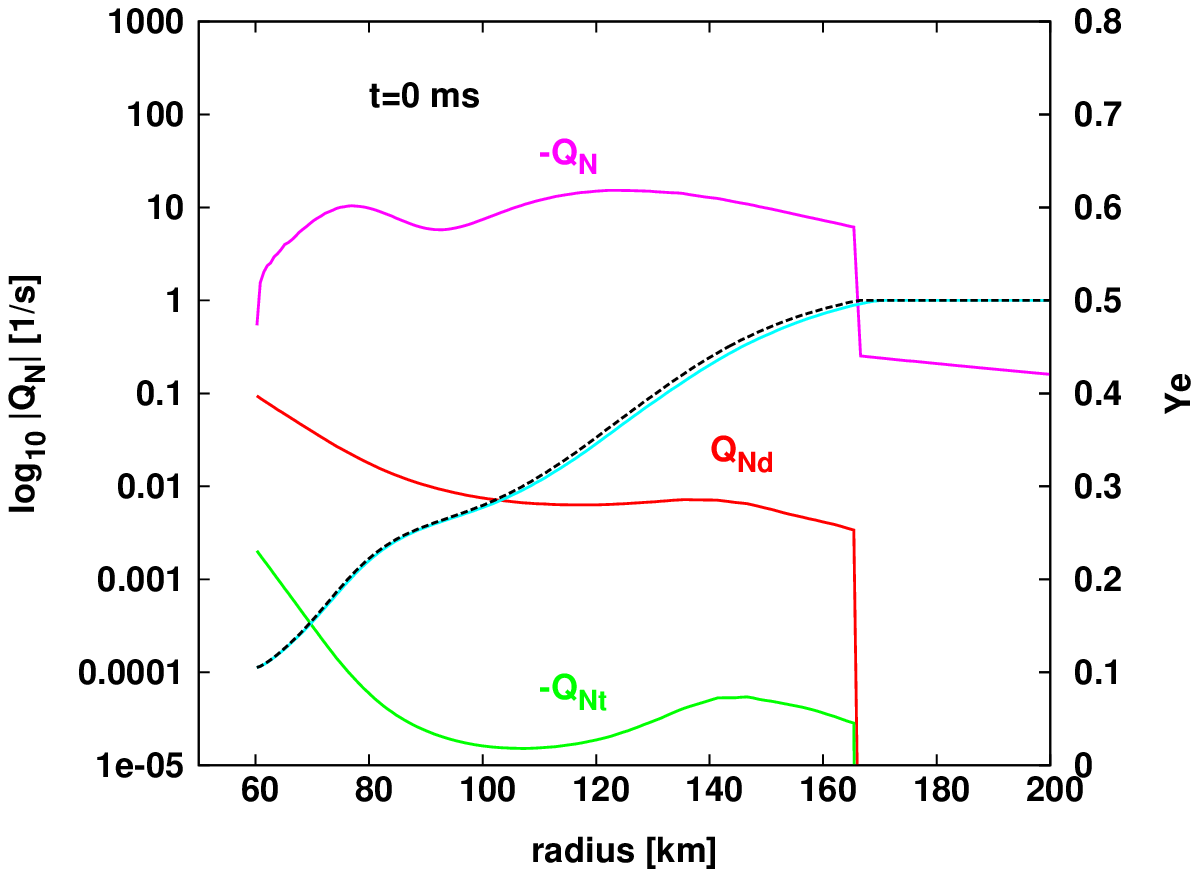}}  \\
  \resizebox{140mm}{!}{\plotone{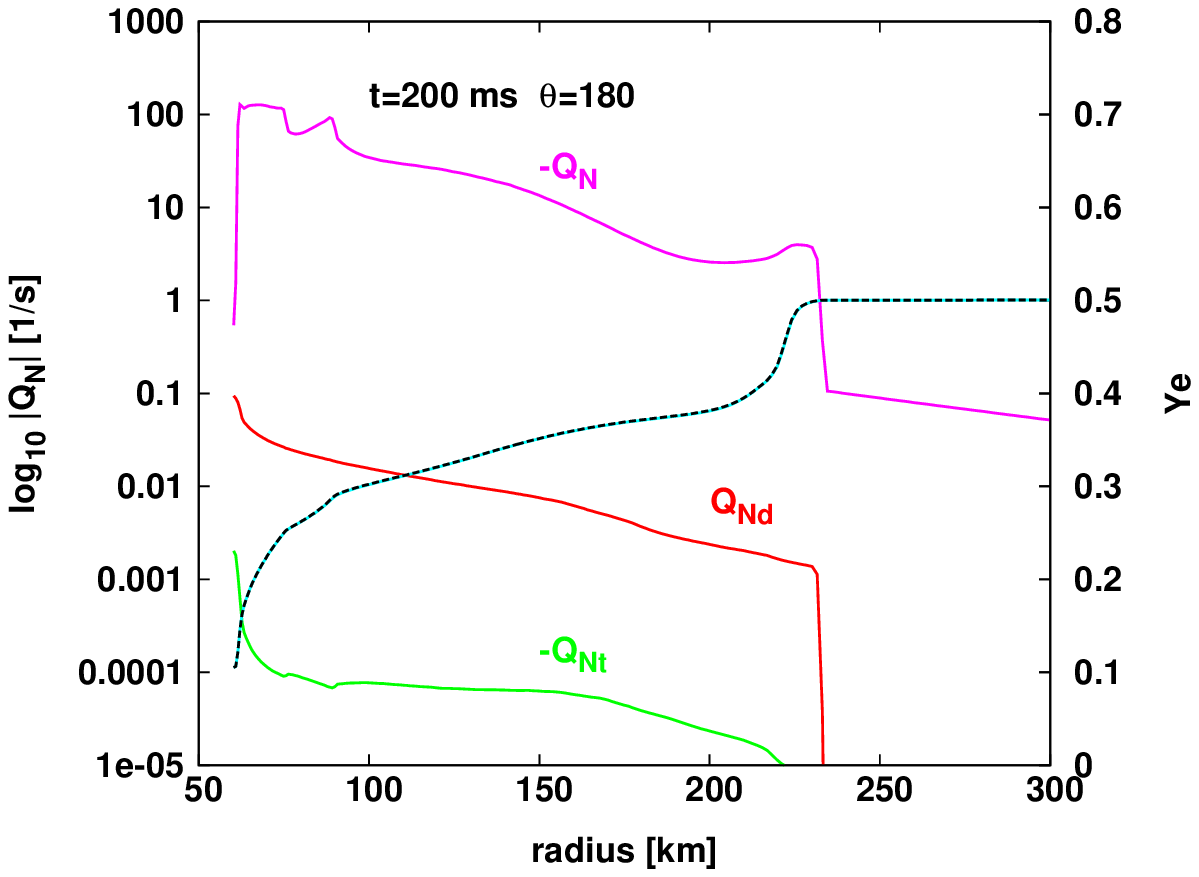}}  
    \end{tabular}
\end{center}
\caption{ The electron fractions (black dotted line) and
 the absolute values of $Q_{N}$ (magenta lines),  $Q_{Nd}$ (red solid lines) and $Q_{Nt}$ (green solid lines) 
at $t=0$ and  200~ms along the radial ray with $\theta=180^{\circ}$.
See the text for the definitions of $Q_N$, $Q_{Nd}$ and $Q_{Nd}$.
 The electron fractions obtained with $Q_{Nd}=Q_{Nt}=0$ are also shown by the cyan solid lines.}
\label{CCcheck}
\end{figure}

The numerical simulations in this paper are admittedly of experimental nature and the numbers we have obtained may be subject to change in more realistic simulations.
We need more systematic investigations, varying not only the neutrino luminosity and mass accretion rate but also the neutrino temperature, mass of a central object and initial perturbation.      
The cooling reactions of light nuclei  such as $d+e^{-} \rightarrow n+n+\nu_e$ 
and $n+p \rightarrow d+\nu+\bar{\nu}$ 
should be incorporated in the calculations,
since deuterons are abundant in the cooling regions as shown in Figs.~\ref{fraheast} and~\ref{frahea52}.
 Recently, \citet{Nasu13} demonstrated that the existence of deuteron reduces the neutrino emissions 
by the electron- and positron captures, 
since the capture rate of deuterons is lower than that of nucleons. 
They made a comparison of the total $e^{\pm}$-capture rates between the models with and without deuterons 
 at  150 ~ms after core bounce, employing the compositions of light nuclei calculated in \citet{Sumiyoshi08}. 
Figs. 1 and 2 in \citet{Nasu13} indicate that this effect becomes remarkable inside the neutrino sphere
but the reduction factor goes down below a few percent near the gain radius ($r \sim$ 90 km).   
Note that this effect is included in our models except for the $e^{\pm}$-captures on deuterons.
Although the latter rates are not available for us at present, their contributions to the cooling
would be less than a few percent of the cooling rates of nucleons in the gain region
and the discussion in this article would not be changed significantly. 
Of course the reactions will become important at the bottom of the cooling region. 
The neutrino transport should be improved so that the neutrino emission from accreting matter could be properly treated.
For more realistic simulations the central part of the core should be also solved self-consistently. 
Last but not least the difference between 2D and 3D should be made clear.
These issues are currently being tackled and will be reported elsewhere.

\acknowledgments 
S. F. thanks T. Sato, S. Nasu, T. Fischer,  W. I. Nakano and Y. Yamamoto for their useful discussions.
S.~F. is supported by the Japan Society for the Promotion of Science Research Fellowship for Young Scientists.
This work is partially supported by the Grant-in-Aid for Scientific Research on Innovative Areas (Nos. 20105004, 20105005),
 the Grant-in-Aid for the Scientific Research (Nos. 22540296, 24103006, 24244036, 24740165),
 the HPCI Strategic Program and Excellent Graduate Schools from the Ministry of Education, Culture, Sports, Science and Technology (MEXT) 
A part of the numerical calculations were carried out on SR16000 at YITP in Kyoto University. 
%
\bibliography{literat}
\newpage
\begin{table}[t]
\begin{center}   
\begin{tabular}{cccccc}
\hline \hline
  Flavor\ \ \ \ &\ \ \ \ \ $n,p$ \ \ \ \ \ &  \ $d$ \ \ \  \ \ \ & \ \ \ \ \ \ $t$ \ \ \ \ \ \ & \ \ \ \ \ \ $h$  \ \ \ \ \ & \ \ \ \ \ \ $\alpha$ \ \ \ \ \\
 \hline
 $\nu_e $ (CC)         & 356.8  & 168.9  & 0.000 & 0.000   & 0.5925  \\
 $\bar{\nu} _e$ (CC) & 557.5  & 169.1  & 8.015   & 0.000     &  1.896  \\
 $\nu_e $ (NC)         & 0.000  &  12.94  & 0.6656 & 0.7446   & 0.008347 \\
 $\bar{\nu}_e$ (NC) & 0.000   & 16.83  & 1.801   & 1.975     &  0.4231 \\
 $\nu_{\mu}$ (NC)   & 0.000   & 139.0  & 31.05   & 32.55    &  17.61   \\
\hline \hline
\end{tabular}
\caption{\label{tab1}
The heating rates of nucleons, deuterons, tritons, helions and alpha particles in unit of  MeV/sec per baryon. 
$X_p=0.5$ and $X_n=0.5$ are assumed in the calculations for nucleons.
The mass fraction of each  light nucleus is set to unity in the calculation for light nuclei; $X_{d,h,t,\alpha}=1$. 
Other parameters are set at $r=100$ km, $L_{\nu_e,\bar{\nu}_e}=5.0\times 10^{52}$ erg s$^{-1}$, $L_{\nu_{\mu}}=0.5 \times L_{\nu_e,\bar{\nu}_e}$, 
$T_{\nu_e}=4$ MeV,  $T_{\bar{\nu}_e}=5$ MeV and $T_{\nu_{\mu}}=10$ MeV.}
\end{center}   
\end{table}
%
%
%

\bibliography{literat}

\begin{thebibliography}{} 
\bibitem[Arcones et~al.(2008)]{Arcones08}{Arcones}, A., {Mart{\'{\i}}nez-Pinedo}, G., {O'Connor}, E., {Schwenk}, A., {Janka}, H.-T., {Horowitz}, C.~J., \& {Langanke}, K. 2008, \prc, 78, 015806
\bibitem[Barnea et~al.(2008)]{Barnea08}Barnea, N. \& Gazit, D. 2008, Few-Body Systems, 43, 5
\bibitem[Blinnikov et al.(2011)] {Blinnikov11} Blinnikov, S. I., Panov, I. V., Rudzsky, M. A. \& Sumiyoshi, K. 2011, A\&A, 535, A37
\bibitem[Blondin et al.(2003)]{Blondin03}Blondin, J. M., Mezzacappa, A. \& DeMarino, C., 2003, \apj, 584, 971
\bibitem[Botvian et al.(2004)]{Botvina04} Botvina, A.S. \& Mishustin, I.N. 2004 Phys. Lett. B 584, 233.
\bibitem[Botvina et al.(2010)]{Botvina10} Botvina, A.S. \& Mishustin, I.N. 2010 Nucl. Phys.  A 843, 98. 
\bibitem[Bruenn et al.(2013)]{Bruenn13} {{Bruenn}, S.~W. and {Mezzacappa}, A. and {Hix}, W.~R. and {Lentz}, E.~J. and 
	{Bronson Messer}, O.~E. and {Lingerfelt}, E.~J. and {Blondin}, J.~M. \&	{Endeve}, E. and {Marronetti}, P. and {Yakunin}, K.~N.} 2013, \apj 767, 6
\bibitem[Buras et al.(2006a)]{Buras06a}Buras, R., Rampp, M., Janka, H.-Th. \& Kifonidis, K., 2006a, \aap, 447, 1049
\bibitem[Buras et al.(2006b)]{Buras06b}Buras, R., Janka, H.-Th., Rampp, M. \& Kifonidis, K., 2006b, \aap, 457, 281
\bibitem[Burrows et al.(1995)]{Burrows95}Burrows, A., Hayes, J. \& Fryxell, B. A., 1995, \apj, 450, 830
\bibitem[Burrows (2012)] {Burrows12} Burrows, A.  2012 arXiv:astro-ph 1210.4921

\bibitem[Couch (2013)] {Couch13} Couch, S.~M. 2013, \apj, 765, 29
\bibitem[Fern{\'a}ndez \& {Thompson}(2009a)]{Fernandez09a} {Fern{\'a}ndez}, R. \& {Thompson}, C. 2009{\natexlab{a}}, \apj, 697, 1827
\bibitem[Fern{\'a}ndez \& {Thompson}(2009b)]{Fernandez09b} {Fern{\'a}ndez}, R. \& {Thompson}, C. 2009{\natexlab{b}}, \apj, 703, 1464
\bibitem[Fern{\'a}ndez(2010)]{Fernandez10} Fern{\'a}ndez, R.\ 2010, \apj, 725, 1563 
\bibitem[Foglizzo et al.(2006)]{Foglizzo06}Foglizzo, T., Scheck, L. \& Janka, H.-Th., 2006, \apj, 652, 1436
\bibitem[Foglizzo et al.(2007)]{Foglizzo07} Foglizzo, T., Galletti, P., Scheck, L., \& Janka, H.-T.\ 2007, \apj, 654, 1006 
\bibitem[Foglizzo et al.(2012)]{Foglizzo12} Foglizzo, T., Masset, F., Guilet, J., \& Durand, G.\ 2012, Physical Review Letters, 108, 051103
\bibitem[{Fryer} {et~al.}(2002)]{Fryer02} {Fryer}, C.~L., {Holz}, D.~E., \& {Hughes}, S.~A. 2002, \apj, 565, 430
\bibitem[{{Fryer}(2004)}]{Fryer04}{Fryer}, C.~L. 2004, Astrophys. J. Lett., 601, L175
\bibitem[Furusawa et al.(2011)]{Furusawa11} {Furusawa}, S., {Yamada}, S., {Sumiyoshi}, K. \& {Suzuki}, H. 2011 \apj, 738,  178. 
\bibitem[Furusawa et al.(2013)]{Furusawa13} {Furusawa}, S., {Sumiyoshi}, K.,  {Yamada} S. \& {Suzuki}, H. 2013, arXiv:astro-ph 1305.1508 
\bibitem[Gazit et al.(2004)]{Gazit04} {Gazit}, D. and {Barnea}, N.,2004  \prc, 70, 8801
\bibitem[Hanke et al.(2012)]{Hanke12} Hanke, F., Marek, A., Mueller, B., \& Janka, H.-T.\ 2012, \apj 755, 138 
\bibitem[Hanke et al.(2013)]{Hanke13} {Hanke}, F. and {Marek}, A. and {M{\"u}ller}, B. and {Janka}, H.-T. 
arXiv:astro-ph 1303.6269
\bibitem[Haxton(1988)]{Haxton88} Haxton, W.~C.\ 1988, Physical Review Letters, 60, 1999
\bibitem[Hempel et al.(2010)]{Hempel10}   Hempel, M. \& Schaffner-Bielich, J. 2010  Nucl. Phys. A 837, 210
\bibitem[Hempel et al.(2012)]{Hempel12} Hempel, M., Fischer, T., Schaffner-Bielich, J. \& Liebend{\"o}rfer, M. 2012, \apj, 748, 70.
\bibitem[Herant et al.(1992)]{Herant92}Herant, M., Benz, W. \& Colgate, S., 1992, \apj, 395, 642
%
\bibitem[Iwakami et al.(2008)]{Iwakami08}{Iwakami}, W., {Kotake}, K., {Ohnishi}, N., {Yamada}, S., \& {Sawada}, K. 2008, \apj, 678, 1207
\bibitem[Iwakami et al.(2009)]{Iwakami09}{Iwakami}, W., {Kotake}, K., {Ohnishi}, N., {Yamada}, S., \& {Sawada}, K. 2009, \apj, 700, 232   
\bibitem[Janka (2012)] {Janka12}Janka, H.-T, 2012, ARNPS. 62, 407
\bibitem[Kotake et~al. (2006)]{Kotake06} Kotake, K., Sato, K., Takahashi, K., 2006, Reports of Progress in Physics, 69, 971
\bibitem[Kuroda et al.(2012)]{Kuroda12} Kuroda, T., Kotake, K., \& Takiwaki, T.\ 2012, \apj, 755, 11
\bibitem[Lattimer et al.(1991)]{Lattimer91} Lattimer, J. M. \& Swesty, F. D. 1991,  Nucl. Phys. A, 535, 331
\bibitem[Langanke et al.(2008)]{Langanke08} Langanke, K. \& Martinez-Pinedo, G., Mu\"uller, B., Janka, H.-Th., Marek, A., Hix, W. R., Juodagalvis, A. \& Sampaio J. M.
2008, \prl 100, 011101
\bibitem[Marek et al.(2009)]{Marek09a}Marek, A. \& Janka, H.-Th., 2009, \apj, 694, 664
\bibitem[Marek et al.(2009)]{Marek09b} Marek, A., Janka, H.-T., M\"uller, E.\ 2009, \aap, 496, 475 
\bibitem[M\"uller et al.(2012)]{Muller12} Mueller, B., Janka, H.-T., \& Heger, A.\ 2012, \apj, 761, 72
\bibitem[Murphy et al.(2013)]{Murphy13}{{Murphy}, J.~W. and {Dolence}, J.~C. \& {Burrows}, A.} \apj, 771, 52 
\bibitem[Nagakura et al.(2008)]{Nagakura08} Nagakura, H., \& Yamada, S.\ 2008, \apj, 689, 391
\bibitem[Nagakura et al.(2011)]{Nagakura11} Nagakura, H., Ito, H., Kiuchi, K., \& Yamada, S.\ 2011, \apj, 731, 80 
\bibitem[Nagakura et al.(2013)]{Nagakura13} Nagakura, H., Yamamoto, Y., \& Yamada, S.\ 2013, \apj, 765, 123
\bibitem[Nakamura et al.(2009)]{Nakamura09} Nakamura, S. X., Sumiyoshi, K. \& Sato, T. 2009, \prc 80, 035802 
\bibitem[Nakamura et al.(2012)]{Nakamura12} Nakamura, K., Takiwaki, T., Kotake, K., Nishimura, N.  arXiv:astro-ph 1207.5955
\bibitem[Nasu et al. (2013)]{Nasu13}Nasu, S., Sato, T., Nakamura, S.~X., {Sumiyoshi}, K. {Myhrer}, F. and Kubodera, K.,2013, Few-Body Systems, 54, 1595
\bibitem[O'Connor et al.(2007)]{OConnor07} O'Connor, E., Gazit, D., Horowitz, C.J., Schwenk, A., Barnea, N., 2007 Phys. Rev. C 75, 055803
\bibitem[Ohnishi et al.(2006)]{Ohnishi06}Ohnishi, N., Kotake, K. \& Yamada, S. 2006, \apj, 641, 1018
\bibitem[Ohnishi et al.(2007)]{Ohnishi07}Ohnishi, N., Kotake, K. \& Yamada, S. 2007, \apj, 667, 375
\bibitem[Ott et al.(2013)]{Ott13}
{{Ott}, C.~D. and {Abdikamalov}, E. and {M{\"o}sta}, P. and {Haas}, R. and 
	{Drasco}, S. and {O'Connor}, E.~P. and {Reisswig}, C. and {Meakin}, C.~A. \& 
	{Schnetter}, E.} 2013 \apj, 768, 1150
\bibitem[Scheck et al.(2004)]{Scheck04}Scheck, L., Plewa, T., Janka, H.-T., Kifonidis, K., \& M\"uller, E., 2004, \prl, 92, 011103
\bibitem[Scheck et al.(2006)]{Scheck06}{Scheck}, L., {Kifonidis}, K., {Janka}, H., \& {M{\"u}ller}, E. 2006, \aap, 457, 963
\bibitem[Shen et al.(1998a)]{Shen98a} Shen, H., Toki, H., Oyamatsu, K. \& Sumiyoshi, K. 1998a, Nucl. Phys. A, 637, 435
\bibitem[Shen et al.(1998b)]{Shen98b} Shen, H., Toki, H., Oyamatsu, K. \& Sumiyoshi, K. 1998b, Prog. Theor. Phys., 100, 1013
\bibitem[Shen et al.(2011)]{Shen11} Shen, H., Toki, H., Oyamatsu, K. \& Sumiyoshi, K. 2011, Astrophys. J. Supplement 197, 20
\bibitem[G. Shen et al.(2011)]{G.Shen11} Shen G., Horowitz C. J. \& Teige S., 2011, Phys. Rev. C, 83, 035802  
\bibitem[Sumiyoshi et al.(2005)]{Sumiyoshi05}  {Sumiyoshi}, K. and {Yamada}, S. and {Suzuki}, H. and {Shen}, H. and 
	{Chiba}, S. \& {Toki}, H., 2005, \apj, 629, 922
\bibitem[Sumiyoshi et al.(2008)]{Sumiyoshi08} Sumiyoshi, K. \& R{\"o}pke, G. 2008 Phys. Rev. C 77 055804.
\bibitem[Suwa et al.(2010)]{Suwa10}Suwa, Y., Kotake, K., Takiwaki, T., Whitehouse, S. C., Liebend\"orfer, M. \& Sato, K., 2010, \pasj, 62, L49
\bibitem[Suwa et al.(2011)]{Suwa11} Suwa, Y., Kotake, K., Takiwaki, T., Liebend{\"o}rfer, M., \& Sato, K.\ 2011, \apj, 738, 165 
\bibitem[Suwa et al.(2013)]{Suwa13} Suwa, Y., Takiwaki, T., Kotake, K., Fischer, T. Liebend{\"o}rfer, M. \& {Sato}, K. 2013, \apj, 764, 99 
\bibitem[Takiwaki et al.(2012)]{Takiwaki12} Takiwaki, T., Kotake, K. \&  Suwa, Y., 2012, \apj, 749, 98
\bibitem[Yamasaki et al.(2005)]{Yamasaki05} Yamasaki, T., \& Yamada, S.\ 2005, \apj, 623, 1000 
\bibitem[Yamamoto et al.(2013)]{Yamamoto13} Yamamoto, Y.,  Fujimoto, S., Nagakura, H., \& Yamada, S. 2013 \apj,    771, 27 
\end{thebibliography}
\end{document}